\documentclass[aps,prb,reprint,superscriptaddress]{revtex4-2}
\usepackage{graphicx}
\usepackage{amsmath}
\usepackage{amssymb}
\usepackage{color}
\usepackage{hyperref}
\usepackage{changes}

\begin{document}

\title{Colossal magnetoresistance in narrow-gap magnetic semiconductor Eu$_2$CdAs$_2$}

\author{Zhiyu Zhou}
\altaffiliation{Contributed equally to this work}
\affiliation{Key Laboratory of Quantum Materials and Devices of Ministry of Education, School of Physics, Southeast University, Nanjing 211189, China}
\author{Zhi-Cheng Wang}
\altaffiliation{Contributed equally to this work}
\email{wzc@seu.edu.cn}
\affiliation{Key Laboratory of Quantum Materials and Devices of Ministry of Education, School of Physics, Southeast University, Nanjing 211189, China}
\author{Xiyu Chen}
\affiliation{Key Laboratory of Quantum Materials and Devices of Ministry of Education, School of Physics, Southeast University, Nanjing 211189, China}
\author{Xin-Yu Zhao}
\affiliation{School of Physics, Interdisciplinary Center for Quantum Information and State Key Laboratory of Silicon and Advanced Semiconductor Materials, Zhejiang University, Hangzhou 310058, China}
\author{Yaran Duan}
\affiliation{Key Laboratory of Quantum Materials and Devices of Ministry of Education, School of Physics, Southeast University, Nanjing 211189, China}
\author{Shiran Tao}
\affiliation{Key Laboratory of Quantum Materials and Devices of Ministry of Education, School of Physics, Southeast University, Nanjing 211189, China}
\author{Guang-Han Cao}
\affiliation{School of Physics, Interdisciplinary Center for Quantum Information and State Key Laboratory of Silicon and Advanced Semiconductor Materials, Zhejiang University, Hangzhou 310058, China}

\author{Shuai Dong}
\affiliation{Key Laboratory of Quantum Materials and Devices of Ministry of Education, School of Physics, Southeast University, Nanjing 211189, China}

\date{\today}

\begin{abstract}

Discovering antiferromagnetic materials with a sizable colossal magnetoresistance (CMR) effect and precisely tuning their responses remain significant challenges in materials science. This work reports the synthesis, phase transitions, and magnetotransport properties of the narrow-gap Zintl arsenide Eu$_2$CdAs$_2$. The compound undergoes antiferromagnetic ordering at $T_\mathrm{N} = 9$ K and exhibits semiconducting behavior with a low-temperature activation energy of 3.2 meV, achieving a large CMR effect of $-10^4\%$ at 9 T. Below $T_\mathrm{N}$, a distinct butterfly-shaped magnetoresistance hysteresis emerges without any macroscopic ferromagnetic order, revealing a transport anomaly likely dictated by AFM domain dynamics and/or the formation and percolation of localized magnetic polarons. Furthermore, a systematic survey of reported Eu-based Zintl phases uncovers a nearly exponential scaling trend between the maximum CMR magnitude and the low-temperature activation energy. This correlation demonstrates that bandgap engineering serves as an effective strategy for tailoring magnetotransport, establishing Eu-containing Zintl phases as a versatile platform for exploring and manipulating CMR effects.

\end{abstract}
\maketitle

\section{Introduction}

Colossal magnetoresistance (CMR) is a fundamentally vital phenomenon characterized by a reduction in material resistivity by multiple orders of magnitude upon the application of an external magnetic field. This dramatic response renders CMR systems highly promising candidates for magnetic storage, advanced sensors, and spintronics~\cite{APRamirez1997Colossalmagnetoresistance,Elbio2005ComplexityinStronglyCorrelatedElectronicSystems,Li2013Femtosecondswitchingofmagnetismviastrongly,Ohya2024ColossalMagnetoresistiveSwitchingInduced}. Early benchmarking materials exhibiting CMR primarily comprised ferromagnetic (FM) systems, including rare-earth chalcogenides and hexaborides (e.g., EuO and EuB$_6$)~\cite{kasuyaAnomalousTransportPhenomena1968,sullowMagnetotransportLowCarrier2000,pohlitEvidenceFerromagneticClusters2018}, Mn-based pyrochlores (e.g., Tl$_2$Mn$_2$O$_7$)~\cite{shimakawa1996}, and mixed-valent rare-earth perovskite manganites (e.g., La$_{1-x}$Ca$_x$MnO$_3$)~\cite{salamon2001}. More recently, however, the negligible stray fields and enhanced disturbance immunity of antiferromagnetic (AFM) materials have driven a major paradigm shift toward AFM spintronics and quantum information technologies~\cite{baltz2018,han2023,smejkal2018,yan2020}. Consequently, discovering new materials that host significant CMR effects within an AFM ground state has emerged as a frontline research direction.

Among emerging AFM CMR systems, Eu-containing compounds have attracted widespread attention, with notable examples including EuMnSb$_2$~\cite{sun_field-induced_2021,yinMagnetismMagnetotransportProperties2024}, the Eu$_5M_2X_6$ ($M$ = In, Sn; $X$ = As, Sb) family~\cite{Rosa2020Colossalmagnetoresistanceinanonsymmorphic,Ghosh2022Colossalpiezoresistanceinnarrowgap,Day2025Colossalmagnetoresistanceandanisotropic,Balguri2025Twotypesofcolossalmagnetoresistance}, and the Eu$T_2X_2$ ($T$ = Zn, Cd, Mn; $X$ = As, P, Sb) series~\cite{Wang2021ColossalMagnetoresistancewithoutMixed,Usachov2024Magnetismheatcapacityandelectronic,zhangElectronicBandReconstruction2023a,Luo2023Colossalmagnetoresistanceandtopologicalphasetransition,Krebber2023ColossalmagnetoresistanceinEuZn2P2,krebberMagneticFrustrationWeak2025a}. These Eu-based compounds typically share several distinct hallmarks: narrow bandgaps, low carrier densities (typically $<10^{19}$ cm$^{-3}$), and a divalent Eu$^{2+}$ local-moment AFM ground state~\cite{chenRecentAdvancesUnderstanding2024,kliemt2025}. In addition, their crystal structures conform to the valence-precise Zintl concept, where the anionic sublattice forms covalent bonds obeying the octet rule. Crucially, accumulating experimental evidence reveals that the formation and subsequent percolation of magnetic polarons serve as the common microscopic driver for CMR in these systems, wherein low-density free carriers self-trap into FM clusters around the localized Eu$^{2+}$ moments~\cite{Balguri2025Twotypesofcolossalmagnetoresistance,koppRobustMagneticPolaron2026,Cook2025MagneticpolaronformationinEuZn2P2,liColossalMagnetoresistanceSpinpolarized2024,alecrivilleroMagneticElectronicProperties2023,sunkoSpincarrierCouplingInduced2023,sarTopologicallyTrivialSemiconducting2026}.

Despite the growing library of individual Eu-based CMR materials and the apparent commonality in their underlying physics, a comprehensive comparative analysis across different Zintl families remains lacking. Consequently, a systematic methodology to predict new candidates or rationally manipulate their CMR magnitudes has yet to be established.

In this work, we report a colossal CMR effect in the narrow-gap AFM Zintl semiconductor Eu$_2$CdAs$_2$, which features a low-temperature activation energy of $E_\mathrm{a} = 3.2$ meV and undergoes an AFM transition at $T_\mathrm{N} = 9$ K~\cite{Wang2011SynthesisCrystalandElectronicStructures}. The material exhibits a maximum CMR magnitude reaching $-10^4\%$ at 9 T near $T_\mathrm{N}$. Below the ordering temperature, a characteristic ``butterfly-shaped" magnetoresistance (MR) hysteresis emerges, stemming from the freezing of magnetic polarons within the AFM matrix and/or the AFM domain dynamics. Crucially, going beyond the properties of Eu$_2$CdAs$_2$ itself, a systematic survey across diverse Eu-based Zintl systems uncovers a nearly exponential scaling trend between the maximum CMR magnitude and the low-temperature activation energy. This correlation provides a powerful framework for designing novel Eu-based Zintl phases and tailoring their magnetotransport responses, thereby accelerating the material discovery pipeline and opening new avenues for device optimization via bandgap engineering.


\section{Experimental Section}
\subsection{Synthesis}

Single crystals of Eu$_2$CdAs$_2$ were grown via a Pb-flux method. High-purity Eu pieces ($99.999\%$), Cd shots ($99.99\%$), and As lumps ($99.99\%$) were used as starting materials, with Pb shots ($99.999\%$) serving as the flux. To ensure purity, the Eu pieces were meticulously cleaned using cutting pliers prior to use. All reagents and products were handled inside an argon-filled glove box to prevent oxidation or hydrolysis. The synthesis followed a two-step procedure. First, the constituent elements were roughly mixed in a stoichiometric ratio of $\mathrm{Eu}:\mathrm{Cd}:\mathrm{As} = 2:1:2$ and loaded into an alumina crucible, which was subsequently sealed in an evacuated silica tube. The assembly was heated to 800 $^\circ$C over 10 h and dwelt at this temperature for 2 days to synthesize the polycrystalline precursor. This intermediate product was then ground into fine powder and mixed with Pb shots in a $1:15$ molar ratio. The mixture was again placed in an alumina crucible and sealed in an evacuated silica tube. The tube was heated to 1020 $^\circ$C over 20 h, maintained for another 2 days, and then slowly cooled to 550 $^\circ$C at a rate of 3 $^\circ$C/h. Finally, the excess molten Pb flux was removed by centrifugation. This process yielded high-quality, bar-shaped crystals of Eu$_2$CdAs$_2$. Notably, alternative attempts using Sn or Cd as fluxes were unsuccessful, as they primarily resulted in the formation of EuCd$_2$As$_2$. The obtained Eu$_2$CdAs$_2$ crystals are stable in air.

\subsection{Structure determination}

Suitable crystals were selected for SCXRD analysis. Data collections were carried out at 150 K using a Bruker D8 Venture diffractometer, which was equipped with an I$\mu$S 3.0 Dual Wavelength system (Mo $K\alpha$ radiation, $\lambda$ = 0.71073 \AA) and an APEX-II CCD detector. The collected frames were reduced and corrected using the Bruker SAINT software suite. The crystal structures were determined using the intrinsic phasing method implemented in the SHELXT structure solution program within the Olex2 environment~\cite{26}. Subsequently, anisotropic refinements were performed with the SHELXL refinement package utilizing the least-squares method~\cite{27}. Complementary x-ray diffraction measurements were conducted at room temperature using a SmartLab diffractometer (Cu $K\alpha$ radiation) to verify the crystal quality. Finally, the crystal structures were visualized using the VESTA software~\cite{28}.

\subsection{Magnetization and heat capacity measurements}

DC magnetization measurements, including ZFC and FC protocols, were performed using a Magnetic Property Measurement System (MPMS3, Quantum Design). Specific heat measurements were carried out in a Physical Property Measurement System (PPMS DynaCool, Quantum Design).
\subsection{Magnetotransport measurements}

Electrical resistivity measurements were performed using a Quantum Design PPMS DynaCool via a standard four-probe technique. Electrical contacts were established using gold wires attached with silver paste. High-quality single crystals were selected and polished into a rectangular bar shape to ensure accurate determination of the geometric factor for resistivity calculations. To ensure consistency, the same single-crystal specimen was used for both resistivity and heat capacity measurements.

\section{Results and Discussion}
\subsection{Crystal structure}

\begin{figure}
	\centering
	\includegraphics[width=0.48\textwidth]{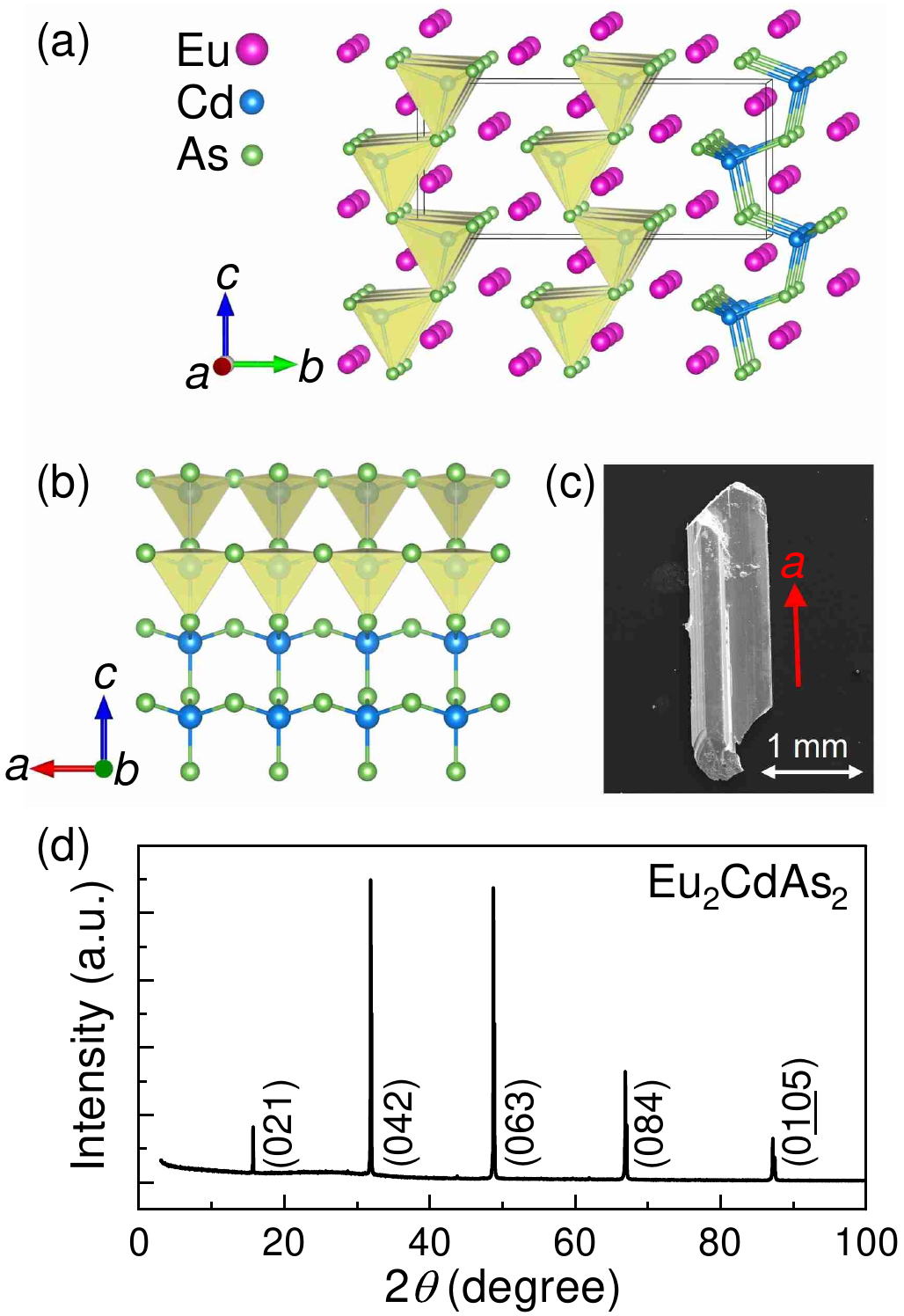}
	\caption{Crystal structure and morphology of Eu$_2$CdAs$_2$.
		(a) Combined polyhedral and ball-and-stick representation of the structure viewed along the crystallographic $a$-axis. The CdAs$_4$ tetrahedra are highlighted in light yellow. Eu, Cd, and As atoms are represented by pink, blue, and green spheres, respectively.
		(b) Structural details of the [CdAs$_2$]$^{4-}$ sheets in Eu$_2$CdAs$_2$, viewed along the $b$-axis.
		(c) Scanning electron microscopy (SEM) micrograph of a representative Eu$_2$CdAs$_2$ single crystal.
		(d) XRD pattern ($\theta-2\theta$ scan) of the Eu$_2$CdAs$_2$ single crystal showing exclusively the (021) family of peaks. The peaks adjacent to the main diffraction peaks are the $K\alpha2$ reflections.}
	\label{F1}
\end{figure}

Eu$_2$CdAs$_2$ is a valence-precise Zintl phase that crystallizes in the Yb$_2$CdSb$_2$-type structure (orthorhombic space group $Cmc2_1$, No. 36). The refined structural parameters are in excellent agreement with previously reported values~\cite{Wang2011SynthesisCrystalandElectronicStructures}, as summarized in Table S1 in the Supplemental Material~\cite{SM}. As illustrated in Fig.~\ref{F1}(a), the framework of Eu$_2$CdAs$_2$ consists of Eu$^{2+}$ cation chains and polyanionic layers composed of corner-sharing CdAs$_4$ tetrahedra. Specifically, the CdAs$_4$ tetrahedra first share corners to form infinite $_\infty^1[\mathrm{CdAs}_3]$ chains along the $a$-axis. These chains are further interconnected along the $c$-axis through additional corner-sharing, with a relative displacement between each pair of adjacent chains (Fig.~\ref{F1}(b)). The resulting quasi-two-dimensional polyanionic slabs, denoted as $_\infty^2[\mathrm{CdAs}_2]^{4-}$, are stacked along the $b$-axis, with Eu$^{2+}$ cations filling the interstitial spaces between the slabs. Figure~\ref{F1}(c) displays a SEM image of a representative Eu$_2$CdAs$_2$ single crystal. Based on the crystal morphology and orientation determined by single-crystal x-ray diffraction (SCXRD, Fig. S1), the elongated direction of the crystal is identified as the crystallographic $a$-axis. Accordingly, we investigate the anisotropic magnetization and resistivity with the magnetic field applied both parallel and perpendicular to this $a$-axis in the following sections. Furthermore, the $\theta-2\theta$ XRD pattern shown in Fig.~\ref{F1}(d) exhibits sharp $(021)$ family diffraction peaks without any detectable impurity phases, demonstrating the high structural quality of our Eu$_2$CdAs$_2$ single crystals.

\subsection{Phase transitions}

\begin{figure*}
	\centering
	\includegraphics[width=0.75\textwidth]{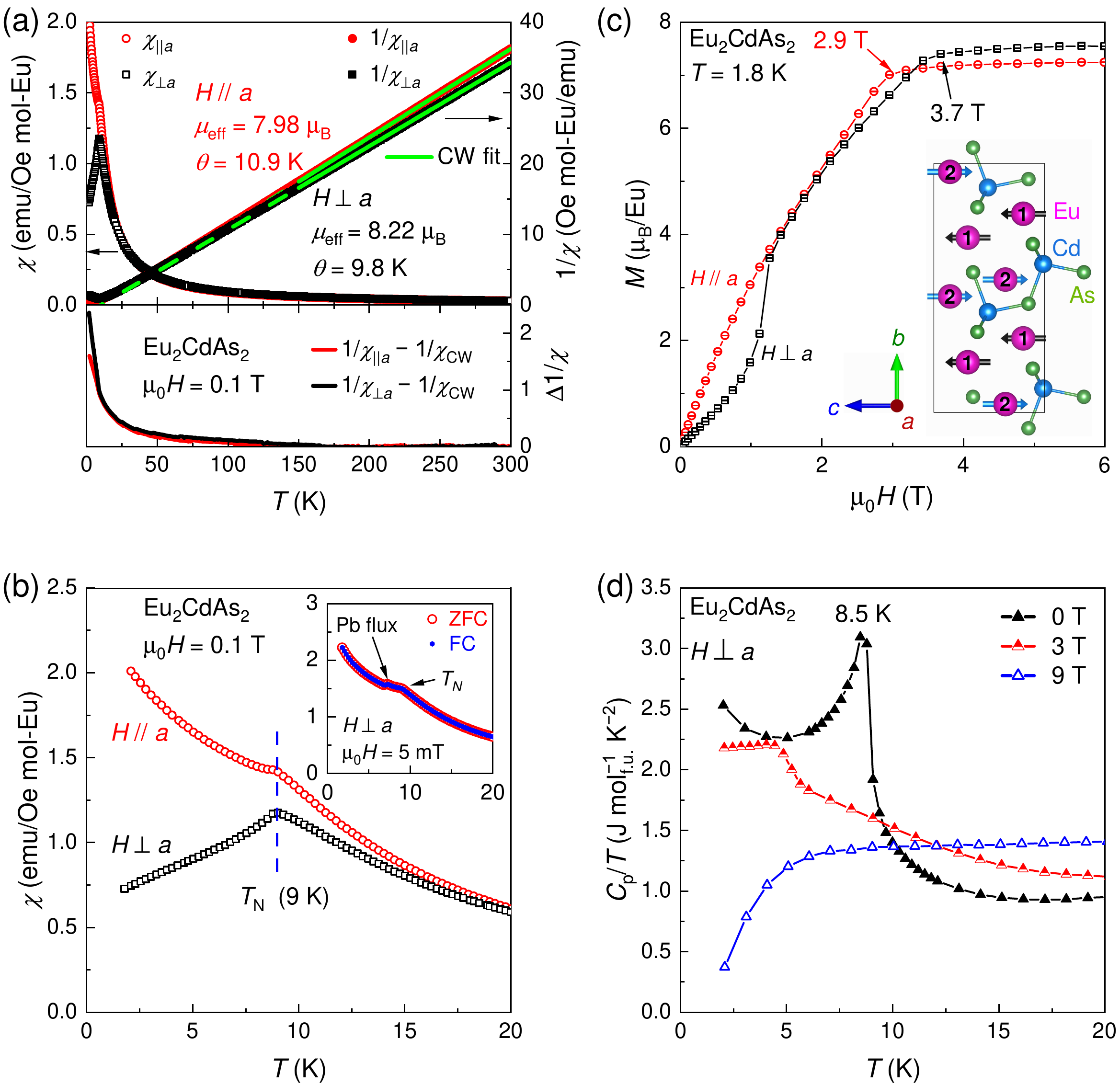}
	\caption{Magnetic phase transition of Eu$_2$CdAs$_2$.
	(a) Temperature dependence of the magnetic susceptibility $\chi(T)$ (left axis of upper panel) measured in an applied field of 0.1 T, and the corresponding Curie-Weiss analysis (right axis of upper panel). Open red circles and open black squares represent data for $H \parallel a$ and $H \perp a$, respectively. The green lines show the Curie-Weiss fit to the inverse susceptibility. The lower panel shows deviations of the $1/\chi$ data from the linear fit.
	(b) Enlarged view of the $\chi(T)$ curves highlighting the magnetic transition at 9 K. The inset compares zero-field-cooling (ZFC, red) and field-cooling (FC, blue) data collected at 5 mT.
	(c) Field-dependent magnetization $M(H)$ for $H \parallel a$ (red circles) and $H \perp a$ (black squares). The schematic in the lower right corner depicts a proposed magnetic structure.
	(d) Temperature dependence of the heat capacity, plotted as $C_\mathrm{p}/T$ versus $T$, under various magnetic fields.}
	\label{F2}
\end{figure*}

The magnetic properties of Eu$_2$CdAs$_2$ are summarized in Fig.~\ref{F2}. Panel (a) presents the temperature dependence of the magnetic susceptibility ($\chi$) and inverse susceptibility ($1/\chi$) measured in an applied field of 0.1 T. The $\chi(T)$ curves exhibit a distinct magnetic transition and weak magnetocrystalline anisotropy above the transition temperature. A magnified view of this transition is provided in Fig.~\ref{F2}(b). Both $\chi_{\parallel a}(T)$ ($H \parallel a$) and $\chi_{\perp a}(T)$ ($H \perp a$) show a transition peak at 9 K, indicating an AFM ground state for Eu$_2$CdAs$_2$. This AFM nature is further supported by the overlapping zero-field-cooled (ZFC) and field-cooled (FC) data collected at 5 mT, as shown in the inset of Fig.~\ref{F2}(b). Notably, while $\chi_{\perp a}(T)$ reaches a maximum at $T_\mathrm{N}$, $\chi_{\parallel a}(T)$ continues to increase below the transition. This behavior suggests that the easy magnetization direction lies perpendicular to the crystallographic $a$-axis, i.e., within the $bc$-plane. At high temperatures, Eu$_2$CdAs$_2$ exhibits typical paramagnetic behavior, which is well-described by the standard Curie-Weiss law, $\chi(T) = C/(T - \theta)$. By fitting the data between 150 and 300 K, we obtained an effective magnetic moment ($\mu_\mathrm{eff}$) of approximately 8 $\mu_\mathrm{B}$ and Weiss temperatures ($\theta$) ranging from 10 to 11 K. The derived $\mu_\mathrm{eff}$ is close to the theoretical value for Eu$^{2+}$, confirming the absence of mixed-valence states. Moreover, the positive Weiss temperatures ($\theta > 0$) suggest that FM correlations are dominant in Eu$_2$CdAs$_2$, despite the eventual AFM spin configuration. As shown in the lower panel of Fig.~\ref{F2}(a), the magnetic susceptibility exhibits a clear deviation from Curie-Weiss behavior below 50 K, well above $T_\mathrm{N}$.

To further characterize the magnetic behavior, $M(H)$ curves were measured at 1.8 K for both $H \perp a$ and $H \parallel a$ (Fig.~\ref{F2}(c)). Consistent with divalent Eu, the magnetization reaches a saturation value of 7 $\mu_\mathrm{B}$ at approximately 3 T for both field directions. A metamagnetic transition is clearly observed in the $M_{\perp a}(H)$ curve; at a critical field of 1.1 T, a sudden change in slope occurs, after which the field dependence of $M_{\perp a}(H)$ parallels that of $M_{\parallel a}(H)$. This behavior is consistent with a spin-flop transition in an antiferromagnet with weak magnetocrystalline anisotropy, further confirming that the easy magnetization axis lies within the $bc$-plane. Although the specific crystallographic $b$ and $c$ directions could not be differentiated due to the irregular crystal morphology, a proposed magnetic structure is illustrated in the inset of Fig.~\ref{F2}(c). In this model, spins are coupled antiferromagnetically within the $bc$-plane. Given the dominant FM correlations inferred from the Curie-Weiss analysis, we propose that the magnetic interactions along the $a$-axis (the Eu chain direction) are likely FM.

Specific heat measurements were employed to corroborate the magnetic phase transition. Figure~\ref{F2}(d) displays the temperature dependence of $C_\mathrm{p}/T$. At zero field, a sharp specific heat anomaly is observed at 8.5 K, which closely aligns with the $T_\mathrm{N} = 9$ K identified from magnetic measurements, marking the onset of long-range AFM order. Interestingly, the specific heat does not decrease monotonically below $T_\mathrm{N}$; instead, it exhibits a low-temperature “upturn” beginning around 5 K. This upturn is suppressed by a 3 T magnetic field, while the AFM transition peak shifts to lower temperatures, evolving into a broad plateau-like feature centered around 4.5 K. At a field of 9 T, both the specific heat peak and the low-temperature upturn disappear. The presence of this upturn at 0 T and its subsequent suppression suggest an additional heat capacity contribution beyond the AFM transition at $T_\mathrm{N}$. In the absence of a magnetic counterpart, this anomaly can be attributed to electronic phase separation, which is driven by the robust formation of magnetic polarons and the resulting interfacial energy fluctuations between phase-separated domains. Furthermore, magnetic entropy analysis of the heat capacity data (presented in Fig. S3, Supplemental Material) confirms the Eu$^{2+}$ valence state in Eu$_2$CdAs$_2$~\cite{SM}.

Finally, it is important to emphasize that no evidence of a FM component was observed in Eu$_2$CdAs$_2$ through either magnetic or specific heat measurements. The AFM nature has been consistently confirmed across multiple Eu$_2$CdAs$_2$ single-crystal samples (see Fig. S2 in the Supplemental Material for additional data)~\cite{SM}. This absence of an FM component appears to contradict the resistance hysteresis observed in transport measurements, a point that will be addressed in the following section.

\subsection{Magnetotransport}

\begin{figure*}
	\centering
	\includegraphics[width=\textwidth]{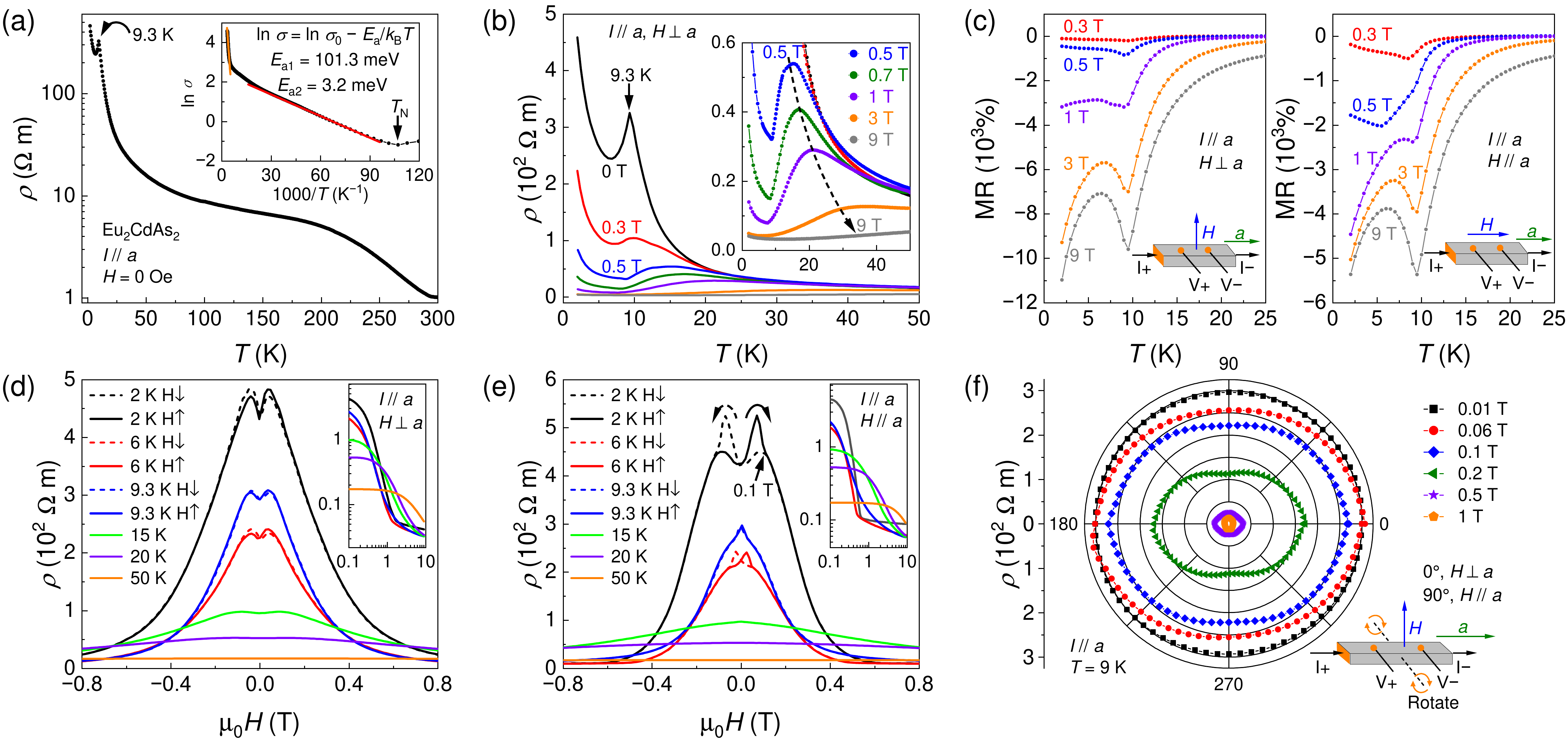}
	\caption{Magnetotransport properties of Eu$_2$CdAs$_2$.
	(a) Temperature dependence of the zero-field resistivity $\rho(T)$ with current $I \parallel a$. The inset shows Arrhenius-type linear fits of $\ln \sigma$ vs. $1000/T$ performed in the high-temperature range ($> 250\text{ K}$) and low-temperature range ($10\text{--}20\text{ K}$), respectively.
	(b) $\rho(T)$ curves under various magnetic fields applied perpendicular to the $a$-axis ($H \perp a$), with $I \parallel a$.
	(c) MR as a function of temperature for $H \perp a$ (left panel) and $H \parallel a$ (right panel).
	(d,e) Field-dependent resistivity $\rho(H)$ at selected temperatures from 2 to 50 K for (d) $H \perp a$ and (e) $H \parallel a$.
	(f) Angular dependence of the resistivity at various magnetic fields, exhibiting modest anisotropy. The angles 0$^\circ$ and 90$^\circ$ correspond to the $H \perp a$ and $H \parallel a$ configurations, respectively.
	}
	\label{F3}
\end{figure*}

Figure~\ref{F3}(a) illustrates the temperature dependence of the zero-field resistivity for Eu$_2$CdAs$_2$ with the current applied along the $a$-axis. The resistivity is plotted on a logarithmic scale. Upon cooling, the resistivity increases by several orders of magnitude, a characteristic feature of semiconducting behavior. The inset displays the Arrhenius plot of $\ln \sigma$ versus $1000/T$, where $\sigma$ represents the electrical conductivity. The plot exhibits distinct linear regimes above $T_\mathrm{N}$. A linear fit of the high-temperature region ($> 250$ K) using the thermal activation model, $\ln \sigma = \ln \sigma_0 - E_\mathrm{a}/(k_\mathrm{B}T)$, yields an activation energy of $E_\mathrm{a1} = 101.3$ meV. This corresponds to an intrinsic bandgap of $202.6$ meV, confirming that $\mathrm{Eu_2CdAs_2}$ is a narrow-gap semiconductor. In the low-temperature regime below 20 K ($1000/T > 50\text{ K}^{-1}$), the linear fit yields a significantly lower activation energy of $E_\mathrm{a2} = 3.2$ meV, where mechanisms such as carrier freeze-out, in-gap states, and localized hopping dominate the charge transport~\cite{Fender2021Narrowgapsemiconductingbehavior}.


Figure~\ref{F3}(b) displays the $\rho(T)$ curves below 50 K under various magnetic fields. At zero field, a prominent peak is observed at 9.3 K, signaling the establishment of long-range AFM order. This transition temperature agrees well with the results from magnetic susceptibility and heat capacity measurements. Below $T_\mathrm{N}$, the resistivity drops sharply, reaching a minimum around 4 K, followed by a secondary upturn at lower temperatures. This upturn can be naturally attributed to activated transport within the remaining semiconducting matrix, which coexists and phase-separates from the conducting regions driven by magnetic polaron formation.


The application of a magnetic field significantly suppresses both the resistivity peak and the low-temperature upturn, leading to a substantial negative MR that spans two orders of magnitude at fields of several Tesla. As the magnetic field increases, the resistivity peak gradually shifts toward higher temperatures and broadens into a hump-like feature, as illustrated in the inset of Fig.~\ref{F3}(b). The resistivity upturn is effectively suppressed at 3 T, resulting in a CMR effect comparable in magnitude to the MR observed at the transition peak. Reorienting the magnetic field from $H \perp a$ (Fig.~\ref{F3}(b)) to $H \parallel a$ (Fig. S4, Supplemental Material) does not significantly alter the magnetotransport behavior of Eu$_2$CdAs$_2$, which is further corroborated by the MR data presented in the panels of Fig.~\ref{F3}(c)~\cite{SM}. Here, MR is defined as $\mathrm{MR} = [(\rho_\mathrm{H} - \rho_0)/\rho_\mathrm{H}] \times 100\%$. This value reaches $-10^3\%$ at 0.5 T for both field orientations and surpasses $-10^4\%$ for $H \perp a$ at 9 T.

The coexistence of peak-type and upturn-type CMR has also been reported in another narrow-gap Zintl phase, Eu$_5$In$_2$As$_6$~\cite{Balguri2025Twotypesofcolossalmagnetoresistance}. These two types of CMR effects in Eu$_5$In$_2$As$_6$ were attributed to polaron physics and charge segregation (or charge ordering), respectively. We propose that these two mechanisms are similarly applicable to Eu$_2$CdAs$_2$ based on the following observations. First, around 50 K (well above $T_\mathrm{N}$), the magnetic susceptibility begins to deviate from Curie-Weiss behavior (lower panel of Fig.~\ref{F2}(a)), which is often regarded as a signature of short-range magnetic correlations within magnetic polarons. Second, a wealth of recent experimental evidence supports magnetic polaron percolation as the microscopic origin of CMR in AFM Eu-based Zintl compounds~\cite{Balguri2025Twotypesofcolossalmagnetoresistance,koppRobustMagneticPolaron2026,Cook2025MagneticpolaronformationinEuZn2P2,liColossalMagnetoresistanceSpinpolarized2024,alecrivilleroMagneticElectronicProperties2023,sunkoSpincarrierCouplingInduced2023,sarTopologicallyTrivialSemiconducting2026}.  Third, regarding the upturn-type CMR in Eu$_2$CdAs$_2$, the external magnetic field strongly suppresses AFM ordering and induces bandgap narrowing or closure via exchange coupling, thereby giving rise to the upturn-type CMR. This mechanism has been previously proposed for Eu$_5$In$_2$As$_6$ and supported by theoretical studies~\cite{Balguri2025Twotypesofcolossalmagnetoresistance,santos-cottin2023}.

%
%

Fig.~\ref{F3}(d) and (e) show the field-dependent resistivity data of Eu$_2$CdAs$_2$. For the $\rho(H)$ curves with $H \perp a$, a slight positive MR is observed at low fields when $T \le 15$ K. This positive component is superimposed on the dominant negative MR, resulting in M-shaped $\rho(H)$ curves. Furthermore, a subtle lack of overlap between the field-increasing and field-decreasing branches is discernible in the low-field region for temperatures below $T_\mathrm{N}$. Notably, this resistivity hysteresis becomes significantly more pronounced for $H \parallel a$, leading to a marked asymmetry between the field-ascending and field-descending sweeps, characteristic of ``butterfly-shaped'' MR. Typically, hysteresis in resistivity has a counterpart in magnetic behavior, such as a hysteresis loop in magnetization. However, as previously emphasized, the magnetic ground state of Eu$_2$CdAs$_2$ is AFM, and no discernible magnetic hysteresis or bifurcation between ZFC and FC curves is observed. Consequently, a plausible physical picture involves FM clusters (magnetic polarons) dispersed within the long-range AFM background. This model simultaneously accounts for the hysteretic resistivity and the non-hysteretic bulk magnetization. It is worth noting that the resistivity hysteresis is observed only below $T_\mathrm{N}$ and intensifies with decreasing temperature, suggesting that this phenomenon emerges when magnetic polarons become ``frozen'' or strongly coupled within the AFM ordered phase. However, the hysteresis in the $\rho(H)$ sweeps does not serve as definitive proof of magnetic polarons alone, as alternative mechanisms, such as AFM domain reorientation or domain wall dynamics, could yield a similar transport response. At present, the available data cannot unambiguously disentangle these competing scenarios. Future experiments, including neutron diffraction, magnetic force microscopy, or angle-resolved torque magnetometry, will be essential to directly track domain behavior and isolate the microscopic driver behind the butterfly-shaped MR hysteresis.


Figure~\ref{F3}(f) illustrates the angular dependence of the resistivity at 9 K, where $0^\circ$ and $90^\circ$ correspond to the $H \perp a$ and $H \parallel a$ configurations, respectively. Overall, the MR anisotropy in Eu$_2$CdAs$_2$ is relatively modest. In a magnetic field of 0.2 T, the anisotropy ratio $\rho(0^\circ)/\rho(90^\circ)$ is approximately 1.5; however, this value shifts to 0.75 at $H = 1$ T. This crossover indicates that the resistivity for $H \perp a$ decreases more rapidly than for $H \parallel a$, a trend consistent with the MR data presented in Fig.~\ref{F3}(c). The more pronounced reduction in resistivity for $H \perp a$ can be attributed to the metamagnetic transition occurring near 1 T, as shown in Fig.~\ref{F2}(c).

\subsection{Discussion}

\begin{figure*}
	\centering
	\includegraphics[width=.75\textwidth]{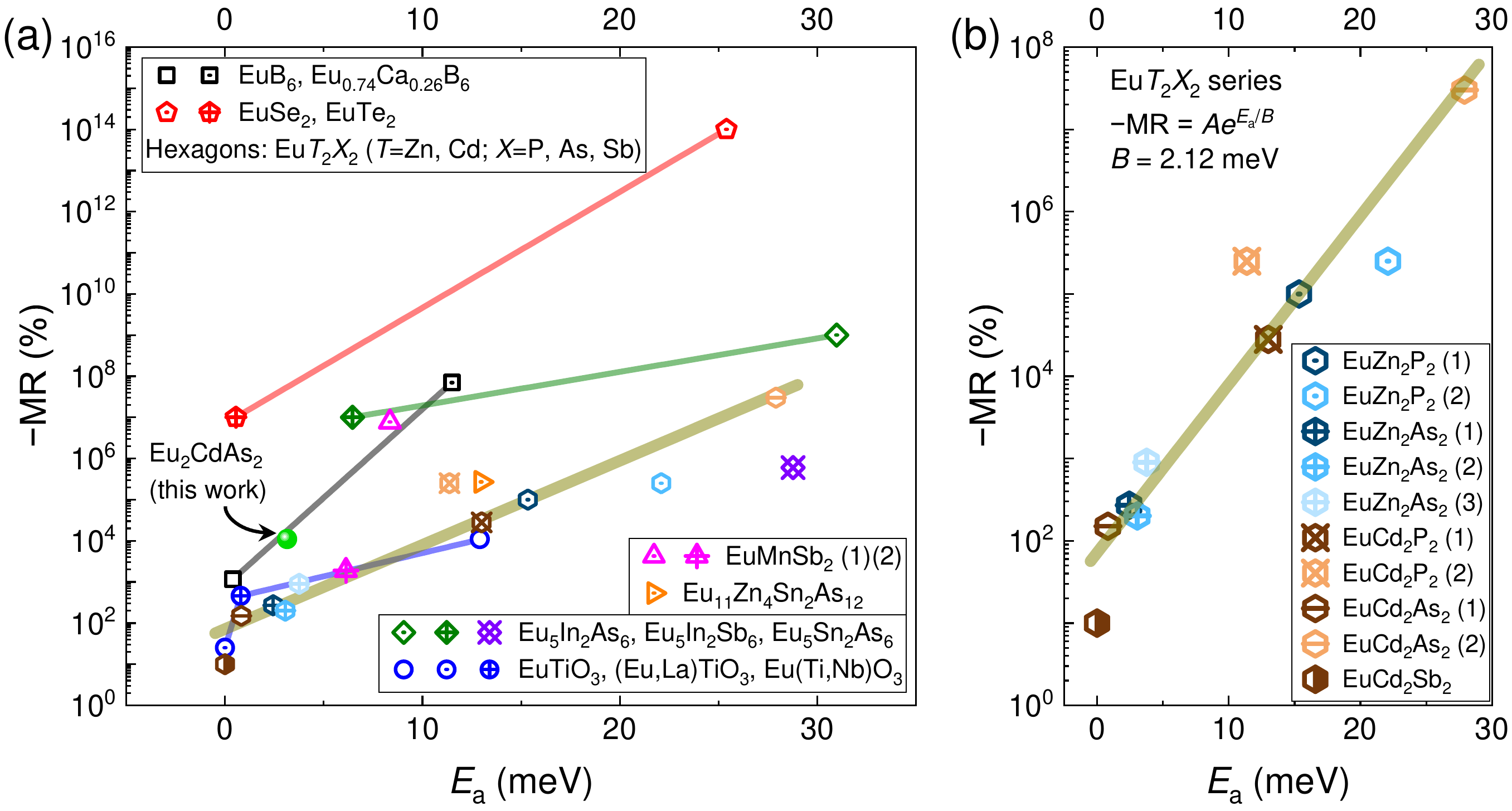}
	\caption{(a) Maximum $-\mathrm{MR}$ of various Eu-containing Zintl compounds (data from single-crystal samples) plotted against their respective low-temperature activation energy. The $-$MR values are presented on a logarithmic scale. Data points belonging to the same structural family are connected by lines, except for the Eu$T_2X_2$ series. Different symbol colors are used for $\mathrm{Eu_5Sn_2As_6}$ and $\mathrm{Eu_5In_2(As,Sb)_6}$ to reflect their structural difference. (b) Exponential fit illustrating the relationship between $-$MR and the activation energy for the Eu$T_2X_2$ ($T$ = Cd, Zn; $X$ = P, As, Sb) series. The brown line represents the linear fit for the data points. All data were compiled from the literature and standardized using the same Arrhenius thermal activation model, with the detailed fitting parameters summarized in Table S2 of Supplemental Material~\cite{sun_field-induced_2021,yi2017,Rosa2020Colossalmagnetoresistanceinanonsymmorphic,Day2025Colossalmagnetoresistanceandanisotropic,Balguri2025Twotypesofcolossalmagnetoresistance,Wang2021ColossalMagnetoresistancewithoutMixed,Usachov2024Magnetismheatcapacityandelectronic,Luo2023Colossalmagnetoresistanceandtopologicalphasetransition,Krebber2023ColossalmagnetoresistanceinEuZn2P2,Cook2025MagneticpolaronformationinEuZn2P2,Iwamoto2022LargeNegativeMagnetoresistance,glushkov2010Enhancementofthecolossalmagnetoresistance,Devlin2018Eu11Zn4Sn2As12,Shi2024AbsenceofWeylnodesinEuCd2As2,Roychowdhury2023AnomalousHallConductivityandNernstEffect,Dong2025Simultaneouscolossalmagnetoresistanceandangular,Yin2020Largenegativemagnetoresistanceintheantiferromagnetic,chen2018Strongelectronphononinteractionandcolossal,Kim2002StructureMagnetismandColossal,Rubi2017Largemagnetoresistanceoverawide,glushkov2007Magneticandtransportproperties,Midya2019Colossalmagnetoresistanceinlow-doped,Wang2022Anisotropyofthemagneticandtransportproperties,Blawat2022UnusualElectricalandMagnetic,SM}.
	}
	\label{F4}
\end{figure*}

The CMR effect in Eu-based materials has long been recognized, tracing back to decades-old reports on Eu-chalcogenides (EuO, EuSe, etc.) and EuB$_6$~\cite{kasuyaAnomalousTransportPhenomena1968,sullowMagnetotransportLowCarrier2000}. In recent years, surging interest in AFM spintronics has fueled the exploration of anomalous transport phenomena in a broad range of Eu-based compounds. Notable examples include the Eu$_5M_2X_6$ ($M$ = In, Sn; $X$ = As, Sb) and Eu$T_2X_2$ ($T$ = Zn, Cd, Mn, In; $X$ = P, As, Sb) families~\cite{Rosa2020Colossalmagnetoresistanceinanonsymmorphic,Ghosh2022Colossalpiezoresistanceinnarrowgap,Day2025Colossalmagnetoresistanceandanisotropic,Balguri2025Twotypesofcolossalmagnetoresistance,Wang2021ColossalMagnetoresistancewithoutMixed,Usachov2024Magnetismheatcapacityandelectronic,zhangElectronicBandReconstruction2023a,Luo2023Colossalmagnetoresistanceandtopologicalphasetransition,Krebber2023ColossalmagnetoresistanceinEuZn2P2,krebberMagneticFrustrationWeak2025a}, all of which feature narrow bandgaps and low carrier densities, much like the present title compound Eu$_2$CdAs$_2$. To contextualize our findings, we surveyed reported transport data for these Eu-based CMR materials and plotted the maximum negative MR ($-\mathrm{MR}$, $y$-axis, logarithmic scale) against the low-temperature activation energy ($E_\mathrm{a}$, $x$-axis, linear scale) in Fig.~\ref{F4}(a). Rather than relying on the intrinsic bandgap from high-temperature data, we chose the low-temperature activation energy because low-temperature transport phenomena (such as CMR) are governed predominantly by extrinsic and complex many-body effects, including electronic phase separation, polaron binding energies, and in-gap defect states. To ensure a consistent comparison, we systematically extracted $E_\mathrm{a}$ for all reference materials by fitting low-temperature data near their respective $-\mathrm{MR}$ maxima. The detailed fitting parameters and temperature ranges are summarized in Supplemental Material~\cite{SM}.

Although the data points in Fig.~\ref{F4}(a), spanning multiple material families, show a wide distribution on the $-\mathrm{MR}$ axis rather than collapsing onto a single universal curve, a clear systematic trend emerges: within each structural family (connected by solid lines), the magnitude of $-\mathrm{MR}$ increases nearly exponentially with increasing $E_\mathrm{a}$. The data dispersion across these Zintl phases is understandable, as structural diversity gives rise to distinct band structures and varying local magnetic interactions that inevitably affect magnetotransport. Furthermore, this scattering is exacerbated by experimental challenges in obtaining precise resistivity measurements for highly resistive samples, which are notoriously sensitive to factors such as contact quality and excitation currents. Overall, this systematic trend points to a unified physical mechanism underlying the CMR effect across these Eu-based Zintl phases.

This correlation becomes even more pronounced when focusing on a single material family. The data points for the isostructural EuZn$_2X_2$ and EuCd$_2X_2$ ($X$ = P, As, Sb) series from Fig.~\ref{F4}(a) are isolated and replotted in Fig.~\ref{F4}(b), and the brown line represents the linear fit. These families serve as an ideal platform to investigate the activation energy dependence of MR due to their identical crystal structures and similar magnetic correlations. Crucially, a vast amount of magnetotransport data has been reported for these compounds recently because their carrier densities are highly sensitive to crystal growth conditions, providing a rich dataset to explore the origin of CMR~\cite{chenRecentAdvancesUnderstanding2024,kliemt2025}. In Fig.~\ref{F4}(b), the linearity between $\ln(-\mathrm{MR})$ and $E_\mathrm{a}$ is remarkably distinct. Fitting the data to the exponential relation $-\mathrm{MR} = A\exp(E_\mathrm{a}/B)$ yields the fitting parameters $A = 10^{1.85}$ and $B = 2.12$ meV. The value of $B$ corresponds to a characteristic temperature of $\sim 24$ K, which is in good agreement with the temperatures at which the maximum MR is experimentally observed in the EuZn$_2X_2$ and EuCd$_2X_2$ families.

The exponential scaling between the MR magnitude and the low-temperature activation energy can be rationalized through the following framework. At zero field, the activated carrier density $n(0) \propto \exp(-E_\mathrm{a}/k_\mathrm{B}T)$ scales the baseline resistivity $\rho(0)$ exponentially with the $E_\mathrm{a}$. Under a magnetic field, the strong $d$-$f$ or $p$-$f$ exchange interactions inherent to Eu-based Zintl phases induce a pronounced band splitting ($\Delta E_{\mathrm{ex}}$) comparable to $E_\mathrm{a}$, closing the effective gap ($E_{\mathrm{eff}} = E_\mathrm{a} - \Delta E_{\mathrm{ex}}$) and triggering an insulator-to-metal transition~\cite{Day2025Colossalmagnetoresistanceandanisotropic,Cook2025MagneticpolaronformationinEuZn2P2,santos-cottin2023,souza2022}. Consequently, the in-field carrier concentration scales as: $$n(H) \propto \exp\left(-\frac{E_\mathrm{a} - \Delta E_{\mathrm{ex}}}{k_\mathrm{B}T}\right) \propto n(0) \cdot \exp\left(\frac{\Delta E_{\mathrm{ex}}}{k_\mathrm{B}T}\right)$$ Here, the $\exp(\Delta E_{\mathrm{ex}}/k_\mathrm{B}T)$ factor drives a magnetic carrier surge that minimizes the high-field resistivity $\rho(H)$. Given that $|\mathrm{MR}| \approx \rho(0)/\rho(H) \gg 1$ and that the field-induced metallic resistivities remain relatively comparable across different materials, the MR magnitude is fundamentally dictated by the exponential growth of the zero-field numerator $\rho(0)$, mathematically inheriting the $\exp(E_\mathrm{a})$ dependence. For isomorphic compounds (e.g., the Eu$T_2X_2$ series), closely matched effective masses, exchange constants, and polaron radii eliminate confounding variables, allowing this pure low-temperature $E_\mathrm{a}$-driven Boltzmann factor to manifest clearly.


Our systematic analysis demonstrates that the low-temperature activation energy largely determines the MR magnitude in Eu-based Zintl materials. The narrow-bandgap characteristic not only renders these compounds highly sensitive to magnetic fields but also makes them remarkably responsive to external tuning. Experimental studies have shown that the activation energy modulation, along with the corresponding variation in MR magnitude, can be achieved through several distinct approaches: (1) Control of crystal growth conditions: Varying the growth parameters of EuZn$_2X_2$ and EuCd$_2X_2$ crystals can alter atomic occupancies and introduce lattice defects, thereby significantly modifying their transport behaviors and CMR effects~\cite{chenRecentAdvancesUnderstanding2024,kliemt2025}; (2) Application of physical pressure or strain: Hydrostatic pressure or lattice strain can continuously compress or expand the crystal lattice, tuning the bandgap value and potentially triggering topological phase transitions~\cite{Gati2021Pressureinducedferromagnetisminthetopological,Luo2023Colossalmagnetoresistanceandtopologicalphasetransition,joseEvolutionMagnetismValence2023}; (3) Chemical substitution: Element replacement exerts a profound influence on both bandgaps and magnetic correlations, as exemplified by the prominent conductivity enhancement via partial substitution of In with Zn in Eu$_5$In$_2$Sb$_6$~\cite{Park2002Eu5In2Sb6Eu5In2xZnxSb6}. Moreover, for potential spintronic applications, electrostatic gating techniques could be utilized to continuously modulate the carrier density and Fermi level, offering an alternative pathway to tune the CMR effect.

\section{Conclusion}

In summary, we have synthesized the Eu-based Zintl phase Eu$_2$CdAs$_2$ and systematically investigated its crystal structure, magnetic properties, and charge transport behaviors. Eu$_2$CdAs$_2$ behaves as a narrow-gap magnetic semiconductor with an intrinsic bandgap of 200 meV and a low temperature activation energy of 3.2 meV. It undergoes an AFM transition at $T_\mathrm{N}$ = 9 K. Crucially, the material exhibits a CMR effect reaching $-10^4\%$ at 9 T. The observation of a prominent ``butterfly-shaped" MR hysteresis for $H \parallel a$ below $T_\mathrm{N}$, coupled with the absence of a macroscopic FM component, suggests that the formation and percolation of localized magnetic polarons may provide a plausible microscopic origin for the CMR effect in Eu$_2$CdAs$_2$.

Beyond the specific properties of Eu$_2$CdAs$_2$, a comprehensive comparative analysis of reported Eu-based Zintl compounds reveals a nearly exponential scaling behavior between the maximum CMR magnitude and the low-temperature activation energy, which becomes exceptionally distinct within isostructural families. This exponential correlation establishes a unifying physical framework that connects seemingly disparate transport phenomena across different Zintl systems. Furthermore, because the energy gap of these materials can be finely modulated via chemical substitution, crystal growth optimization, or external pressure, our findings underscore the immense potential of ``bandgap engineering" to rationally design and tailor desired MR responses.

Consequently, this work not only establishes Eu-based Zintl phases as a fertile model platform for elucidating the intricate interplay among magnetic order, electronic structure, and transport behaviors, but also holds significant promise for functional spintronic applications, including high-sensitivity magnetic sensing, non-volatile magnetic memory, and spintronic logic devices. Given the vast and largely unexplored library of narrow-gap Eu-based Zintl compounds, we anticipate that the design principles established herein will accelerate the discovery of a broader class of materials hosting remarkable CMR effects in the near future.

\begin{acknowledgments}
This work was supported by the National Key Research and Development Program of China (No. 2025YFA1411100) and Young Elite Scientists Sponsorship Program by Jiangsu Association for Science and Technology.
\end{acknowledgments}

\section*{DATA AVAILABILITY}
All data supporting the findings of this study are available from the authors upon reasonable request.

\bibliography{reference}

\begin{thebibliography}{59}%
\makeatletter
\providecommand \@ifxundefined [1]{%
 \@ifx{#1\undefined}
}%
\providecommand \@ifnum [1]{%
 \ifnum #1\expandafter \@firstoftwo
 \else \expandafter \@secondoftwo
 \fi
}%
\providecommand \@ifx [1]{%
 \ifx #1\expandafter \@firstoftwo
 \else \expandafter \@secondoftwo
 \fi
}%
\providecommand \natexlab [1]{#1}%
\providecommand \enquote  [1]{``#1''}%
\providecommand \bibnamefont  [1]{#1}%
\providecommand \bibfnamefont [1]{#1}%
\providecommand \citenamefont [1]{#1}%
\providecommand \href@noop [0]{\@secondoftwo}%
\providecommand \href [0]{\begingroup \@sanitize@url \@href}%
\providecommand \@href[1]{\@@startlink{#1}\@@href}%
\providecommand \@@href[1]{\endgroup#1\@@endlink}%
\providecommand \@sanitize@url [0]{\catcode `\\12\catcode `\$12\catcode
  `\&12\catcode `\#12\catcode `\^12\catcode `\_12\catcode `\%12\relax}%
\providecommand \@@startlink[1]{}%
\providecommand \@@endlink[0]{}%
\providecommand \url  [0]{\begingroup\@sanitize@url \@url }%
\providecommand \@url [1]{\endgroup\@href {#1}{\urlprefix }}%
\providecommand \urlprefix  [0]{URL }%
\providecommand \Eprint [0]{\href }%
\providecommand \doibase [0]{https://doi.org/}%
\providecommand \selectlanguage [0]{\@gobble}%
\providecommand \bibinfo  [0]{\@secondoftwo}%
\providecommand \bibfield  [0]{\@secondoftwo}%
\providecommand \translation [1]{[#1]}%
\providecommand \BibitemOpen [0]{}%
\providecommand \bibitemStop [0]{}%
\providecommand \bibitemNoStop [0]{.\EOS\space}%
\providecommand \EOS [0]{\spacefactor3000\relax}%
\providecommand \BibitemShut  [1]{\csname bibitem#1\endcsname}%
\let\auto@bib@innerbib\@empty
\bibitem [{\citenamefont
  {Ramirez}(1997)}]{APRamirez1997Colossalmagnetoresistance}%
  \BibitemOpen
  \bibfield  {author} {\bibinfo {author} {\bibfnamefont {A.~P.}\ \bibnamefont
  {Ramirez}},\ }\href {https://doi.org/10.1088/0953-8984/9/39/005} {\bibfield
  {journal} {\bibinfo  {journal} {J. Phys. Condens. Matter}\ }\textbf {\bibinfo
  {volume} {9}},\ \bibinfo {pages} {8171} (\bibinfo {year} {1997})}\BibitemShut
  {NoStop}%
\bibitem [{\citenamefont
  {Dagotto}(2005)}]{Elbio2005ComplexityinStronglyCorrelatedElectronicSystems}%
  \BibitemOpen
  \bibfield  {author} {\bibinfo {author} {\bibfnamefont {E.}~\bibnamefont
  {Dagotto}},\ }\href {https://doi.org/10.1126/science.1107559} {\bibfield
  {journal} {\bibinfo  {journal} {Science}\ }\textbf {\bibinfo {volume}
  {309}},\ \bibinfo {pages} {257} (\bibinfo {year} {2005})}\BibitemShut
  {NoStop}%
\bibitem [{\citenamefont {Li}\ \emph {et~al.}(2013)\citenamefont {Li},
  \citenamefont {Patz}, \citenamefont {Mouchliadis}, \citenamefont {Yan},
  \citenamefont {Lograsso}, \citenamefont {Perakis},\ and\ \citenamefont
  {Wang}}]{Li2013Femtosecondswitchingofmagnetismviastrongly}%
  \BibitemOpen
  \bibfield  {author} {\bibinfo {author} {\bibfnamefont {T.}~\bibnamefont
  {Li}}, \bibinfo {author} {\bibfnamefont {A.}~\bibnamefont {Patz}}, \bibinfo
  {author} {\bibfnamefont {L.}~\bibnamefont {Mouchliadis}}, \bibinfo {author}
  {\bibfnamefont {J.}~\bibnamefont {Yan}}, \bibinfo {author} {\bibfnamefont
  {T.~A.}\ \bibnamefont {Lograsso}}, \bibinfo {author} {\bibfnamefont {I.~E.}\
  \bibnamefont {Perakis}},\ and\ \bibinfo {author} {\bibfnamefont
  {J.}~\bibnamefont {Wang}},\ }\href {https://doi.org/10.1038/nature11934}
  {\bibfield  {journal} {\bibinfo  {journal} {Nature}\ }\textbf {\bibinfo
  {volume} {496}},\ \bibinfo {pages} {69} (\bibinfo {year} {2013})}\BibitemShut
  {NoStop}%
\bibitem [{\citenamefont {Ohya}\ \emph {et~al.}(2024)\citenamefont {Ohya},
  \citenamefont {Tsuruoka}, \citenamefont {Kaneda}, \citenamefont {Shinya},
  \citenamefont {Fukushima}, \citenamefont {Takeda}, \citenamefont {Tadano},
  \citenamefont {Endo}, \citenamefont {Anh}, \citenamefont {Masago},
  \citenamefont {Katayama-Yoshida},\ and\ \citenamefont
  {Tanaka}}]{Ohya2024ColossalMagnetoresistiveSwitchingInduced}%
  \BibitemOpen
  \bibfield  {author} {\bibinfo {author} {\bibfnamefont {S.}~\bibnamefont
  {Ohya}}, \bibinfo {author} {\bibfnamefont {S.}~\bibnamefont {Tsuruoka}},
  \bibinfo {author} {\bibfnamefont {M.}~\bibnamefont {Kaneda}}, \bibinfo
  {author} {\bibfnamefont {H.}~\bibnamefont {Shinya}}, \bibinfo {author}
  {\bibfnamefont {T.}~\bibnamefont {Fukushima}}, \bibinfo {author}
  {\bibfnamefont {T.}~\bibnamefont {Takeda}}, \bibinfo {author} {\bibfnamefont
  {Y.}~\bibnamefont {Tadano}}, \bibinfo {author} {\bibfnamefont
  {T.}~\bibnamefont {Endo}}, \bibinfo {author} {\bibfnamefont {L.~D.}\
  \bibnamefont {Anh}}, \bibinfo {author} {\bibfnamefont {A.}~\bibnamefont
  {Masago}}, \bibinfo {author} {\bibfnamefont {H.}~\bibnamefont
  {Katayama-Yoshida}},\ and\ \bibinfo {author} {\bibfnamefont {M.}~\bibnamefont
  {Tanaka}},\ }\href {https://doi.org/https://doi.org/10.1002/adma.202307389}
  {\bibfield  {journal} {\bibinfo  {journal} {Adv. Mater.}\ }\textbf {\bibinfo
  {volume} {36}},\ \bibinfo {pages} {2307389} (\bibinfo {year}
  {2024})}\BibitemShut {NoStop}%
\bibitem [{\citenamefont {Kasuya}\ and\ \citenamefont
  {Yanase}(1968)}]{kasuyaAnomalousTransportPhenomena1968}%
  \BibitemOpen
  \bibfield  {author} {\bibinfo {author} {\bibfnamefont {T.}~\bibnamefont
  {Kasuya}}\ and\ \bibinfo {author} {\bibfnamefont {A.}~\bibnamefont
  {Yanase}},\ }\href {https://doi.org/10.1103/RevModPhys.40.684} {\bibfield
  {journal} {\bibinfo  {journal} {Rev. Mod. Phys.}\ }\textbf {\bibinfo {volume}
  {40}},\ \bibinfo {pages} {684} (\bibinfo {year} {1968})}\BibitemShut
  {NoStop}%
\bibitem [{\citenamefont {S{\"u}llow}\ \emph {et~al.}(2000)\citenamefont
  {S{\"u}llow}, \citenamefont {Prasad}, \citenamefont {Bogdanovich},
  \citenamefont {Aronson}, \citenamefont {Sarrao},\ and\ \citenamefont
  {Fisk}}]{sullowMagnetotransportLowCarrier2000}%
  \BibitemOpen
  \bibfield  {author} {\bibinfo {author} {\bibfnamefont {S.}~\bibnamefont
  {S{\"u}llow}}, \bibinfo {author} {\bibfnamefont {I.}~\bibnamefont {Prasad}},
  \bibinfo {author} {\bibfnamefont {S.}~\bibnamefont {Bogdanovich}}, \bibinfo
  {author} {\bibfnamefont {M.~C.}\ \bibnamefont {Aronson}}, \bibinfo {author}
  {\bibfnamefont {J.~L.}\ \bibnamefont {Sarrao}},\ and\ \bibinfo {author}
  {\bibfnamefont {Z.}~\bibnamefont {Fisk}},\ }\href
  {https://doi.org/10.1063/1.372460} {\bibfield  {journal} {\bibinfo  {journal}
  {J. Appl. Phys.}\ }\textbf {\bibinfo {volume} {87}},\ \bibinfo {pages} {5591}
  (\bibinfo {year} {2000})}\BibitemShut {NoStop}%
\bibitem [{\citenamefont {Pohlit}\ \emph {et~al.}(2018)\citenamefont {Pohlit},
  \citenamefont {R{\"o}{\ss}ler}, \citenamefont {Ohno}, \citenamefont {Ohno},
  \citenamefont {{von Moln{\'a}r}}, \citenamefont {Fisk}, \citenamefont
  {M{\"u}ller},\ and\ \citenamefont
  {Wirth}}]{pohlitEvidenceFerromagneticClusters2018}%
  \BibitemOpen
  \bibfield  {author} {\bibinfo {author} {\bibfnamefont {M.}~\bibnamefont
  {Pohlit}}, \bibinfo {author} {\bibfnamefont {S.}~\bibnamefont
  {R{\"o}{\ss}ler}}, \bibinfo {author} {\bibfnamefont {Y.}~\bibnamefont
  {Ohno}}, \bibinfo {author} {\bibfnamefont {H.}~\bibnamefont {Ohno}}, \bibinfo
  {author} {\bibfnamefont {S.}~\bibnamefont {{von Moln{\'a}r}}}, \bibinfo
  {author} {\bibfnamefont {Z.}~\bibnamefont {Fisk}}, \bibinfo {author}
  {\bibfnamefont {J.}~\bibnamefont {M{\"u}ller}},\ and\ \bibinfo {author}
  {\bibfnamefont {S.}~\bibnamefont {Wirth}},\ }\href
  {https://doi.org/10.1103/PhysRevLett.120.257201} {\bibfield  {journal}
  {\bibinfo  {journal} {Phys. Rev. Lett.}\ }\textbf {\bibinfo {volume} {120}},\
  \bibinfo {pages} {257201} (\bibinfo {year} {2018})}\BibitemShut {NoStop}%
\bibitem [{\citenamefont {Shimakawa}\ \emph {et~al.}(1996)\citenamefont
  {Shimakawa}, \citenamefont {Kubo},\ and\ \citenamefont
  {Manako}}]{shimakawa1996}%
  \BibitemOpen
  \bibfield  {author} {\bibinfo {author} {\bibfnamefont {Y.}~\bibnamefont
  {Shimakawa}}, \bibinfo {author} {\bibfnamefont {Y.}~\bibnamefont {Kubo}},\
  and\ \bibinfo {author} {\bibfnamefont {T.}~\bibnamefont {Manako}},\ }\href
  {https://doi.org/10.1038/379053a0} {\bibfield  {journal} {\bibinfo  {journal}
  {Nature}\ }\textbf {\bibinfo {volume} {379}},\ \bibinfo {pages} {53}
  (\bibinfo {year} {1996})}\BibitemShut {NoStop}%
\bibitem [{\citenamefont {Salamon}\ and\ \citenamefont
  {Jaime}(2001)}]{salamon2001}%
  \BibitemOpen
  \bibfield  {author} {\bibinfo {author} {\bibfnamefont {M.~B.}\ \bibnamefont
  {Salamon}}\ and\ \bibinfo {author} {\bibfnamefont {M.}~\bibnamefont
  {Jaime}},\ }\href {https://doi.org/10.1103/RevModPhys.73.583} {\bibfield
  {journal} {\bibinfo  {journal} {Rev. Mod. Phys.}\ }\textbf {\bibinfo {volume}
  {73}},\ \bibinfo {pages} {583} (\bibinfo {year} {2001})}\BibitemShut
  {NoStop}%
\bibitem [{\citenamefont {Baltz}\ \emph {et~al.}(2018)\citenamefont {Baltz},
  \citenamefont {Manchon}, \citenamefont {Tsoi}, \citenamefont {Moriyama},
  \citenamefont {Ono},\ and\ \citenamefont {Tserkovnyak}}]{baltz2018}%
  \BibitemOpen
  \bibfield  {author} {\bibinfo {author} {\bibfnamefont {V.}~\bibnamefont
  {Baltz}}, \bibinfo {author} {\bibfnamefont {A.}~\bibnamefont {Manchon}},
  \bibinfo {author} {\bibfnamefont {M.}~\bibnamefont {Tsoi}}, \bibinfo {author}
  {\bibfnamefont {T.}~\bibnamefont {Moriyama}}, \bibinfo {author}
  {\bibfnamefont {T.}~\bibnamefont {Ono}},\ and\ \bibinfo {author}
  {\bibfnamefont {Y.}~\bibnamefont {Tserkovnyak}},\ }\href
  {https://doi.org/10.1103/RevModPhys.90.015005} {\bibfield  {journal}
  {\bibinfo  {journal} {Rev. Mod. Phys.}\ }\textbf {\bibinfo {volume} {90}},\
  \bibinfo {pages} {015005} (\bibinfo {year} {2018})}\BibitemShut {NoStop}%
\bibitem [{\citenamefont {Han}\ \emph {et~al.}(2023)\citenamefont {Han},
  \citenamefont {Cheng}, \citenamefont {Liu}, \citenamefont {Ohno},\ and\
  \citenamefont {Fukami}}]{han2023}%
  \BibitemOpen
  \bibfield  {author} {\bibinfo {author} {\bibfnamefont {J.}~\bibnamefont
  {Han}}, \bibinfo {author} {\bibfnamefont {R.}~\bibnamefont {Cheng}}, \bibinfo
  {author} {\bibfnamefont {L.}~\bibnamefont {Liu}}, \bibinfo {author}
  {\bibfnamefont {H.}~\bibnamefont {Ohno}},\ and\ \bibinfo {author}
  {\bibfnamefont {S.}~\bibnamefont {Fukami}},\ }\href
  {https://doi.org/10.1038/s41563-023-01492-6} {\bibfield  {journal} {\bibinfo
  {journal} {Nat. Mater.}\ }\textbf {\bibinfo {volume} {22}},\ \bibinfo {pages}
  {684} (\bibinfo {year} {2023})}\BibitemShut {NoStop}%
\bibitem [{\citenamefont {{\v S}mejkal}\ \emph {et~al.}(2018)\citenamefont {{\v
  S}mejkal}, \citenamefont {Mokrousov}, \citenamefont {Yan},\ and\
  \citenamefont {MacDonald}}]{smejkal2018}%
  \BibitemOpen
  \bibfield  {author} {\bibinfo {author} {\bibfnamefont {L.}~\bibnamefont {{\v
  S}mejkal}}, \bibinfo {author} {\bibfnamefont {Y.}~\bibnamefont {Mokrousov}},
  \bibinfo {author} {\bibfnamefont {B.}~\bibnamefont {Yan}},\ and\ \bibinfo
  {author} {\bibfnamefont {A.~H.}\ \bibnamefont {MacDonald}},\ }\href
  {https://doi.org/10.1038/s41567-018-0064-5} {\bibfield  {journal} {\bibinfo
  {journal} {Nat. Phys.}\ }\textbf {\bibinfo {volume} {14}},\ \bibinfo {pages}
  {242} (\bibinfo {year} {2018})}\BibitemShut {NoStop}%
\bibitem [{\citenamefont {Yan}\ \emph {et~al.}(2020)\citenamefont {Yan},
  \citenamefont {Feng}, \citenamefont {Qin}, \citenamefont {Zhou},
  \citenamefont {Guo}, \citenamefont {Wang}, \citenamefont {Chen},
  \citenamefont {Zhang}, \citenamefont {Wu}, \citenamefont {Jiang},\ and\
  \citenamefont {Liu}}]{yan2020}%
  \BibitemOpen
  \bibfield  {author} {\bibinfo {author} {\bibfnamefont {H.}~\bibnamefont
  {Yan}}, \bibinfo {author} {\bibfnamefont {Z.}~\bibnamefont {Feng}}, \bibinfo
  {author} {\bibfnamefont {P.}~\bibnamefont {Qin}}, \bibinfo {author}
  {\bibfnamefont {X.}~\bibnamefont {Zhou}}, \bibinfo {author} {\bibfnamefont
  {H.}~\bibnamefont {Guo}}, \bibinfo {author} {\bibfnamefont {X.}~\bibnamefont
  {Wang}}, \bibinfo {author} {\bibfnamefont {H.}~\bibnamefont {Chen}}, \bibinfo
  {author} {\bibfnamefont {X.}~\bibnamefont {Zhang}}, \bibinfo {author}
  {\bibfnamefont {H.}~\bibnamefont {Wu}}, \bibinfo {author} {\bibfnamefont
  {C.}~\bibnamefont {Jiang}},\ and\ \bibinfo {author} {\bibfnamefont
  {Z.}~\bibnamefont {Liu}},\ }\href {https://doi.org/10.1002/adma.201905603}
  {\bibfield  {journal} {\bibinfo  {journal} {Adv. Mater.}\ }\textbf {\bibinfo
  {volume} {32}},\ \bibinfo {pages} {1905603} (\bibinfo {year}
  {2020})}\BibitemShut {NoStop}%
\bibitem [{\citenamefont {Sun}\ \emph {et~al.}(2021)\citenamefont {Sun},
  \citenamefont {Wang}, \citenamefont {Mu}, \citenamefont {Wang}, \citenamefont
  {Wang}, \citenamefont {Wu}, \citenamefont {Wang}, \citenamefont {Zhou},\ and\
  \citenamefont {Chen}}]{sun_field-induced_2021}%
  \BibitemOpen
  \bibfield  {author} {\bibinfo {author} {\bibfnamefont {Z.~L.}\ \bibnamefont
  {Sun}}, \bibinfo {author} {\bibfnamefont {A.~F.}\ \bibnamefont {Wang}},
  \bibinfo {author} {\bibfnamefont {H.~M.}\ \bibnamefont {Mu}}, \bibinfo
  {author} {\bibfnamefont {H.~H.}\ \bibnamefont {Wang}}, \bibinfo {author}
  {\bibfnamefont {Z.~F.}\ \bibnamefont {Wang}}, \bibinfo {author}
  {\bibfnamefont {T.}~\bibnamefont {Wu}}, \bibinfo {author} {\bibfnamefont
  {Z.~Y.}\ \bibnamefont {Wang}}, \bibinfo {author} {\bibfnamefont {X.~Y.}\
  \bibnamefont {Zhou}},\ and\ \bibinfo {author} {\bibfnamefont {X.~H.}\
  \bibnamefont {Chen}},\ }\href {https://doi.org/10.1038/s41535-021-00397-4}
  {\bibfield  {journal} {\bibinfo  {journal} {npj Quantum Mater.}\ }\textbf
  {\bibinfo {volume} {6}},\ \bibinfo {pages} {94} (\bibinfo {year}
  {2021})}\BibitemShut {NoStop}%
\bibitem [{\citenamefont {Yin}\ \emph {et~al.}(2024)\citenamefont {Yin},
  \citenamefont {Shi}, \citenamefont {Liu}, \citenamefont {Kan}, \citenamefont
  {Qin}, \citenamefont {Feng}, \citenamefont {He}, \citenamefont {Cao},
  \citenamefont {Xu}, \citenamefont {Ling}, \citenamefont {Tong}, \citenamefont
  {Pi},\ and\ \citenamefont
  {Han}}]{yinMagnetismMagnetotransportProperties2024}%
  \BibitemOpen
  \bibfield  {author} {\bibinfo {author} {\bibfnamefont {H.}~\bibnamefont
  {Yin}}, \bibinfo {author} {\bibfnamefont {X.}~\bibnamefont {Shi}}, \bibinfo
  {author} {\bibfnamefont {X.}~\bibnamefont {Liu}}, \bibinfo {author}
  {\bibfnamefont {X.}~\bibnamefont {Kan}}, \bibinfo {author} {\bibfnamefont
  {Y.}~\bibnamefont {Qin}}, \bibinfo {author} {\bibfnamefont {Q.}~\bibnamefont
  {Feng}}, \bibinfo {author} {\bibfnamefont {M.}~\bibnamefont {He}}, \bibinfo
  {author} {\bibfnamefont {L.}~\bibnamefont {Cao}}, \bibinfo {author}
  {\bibfnamefont {H.}~\bibnamefont {Xu}}, \bibinfo {author} {\bibfnamefont
  {L.}~\bibnamefont {Ling}}, \bibinfo {author} {\bibfnamefont {W.}~\bibnamefont
  {Tong}}, \bibinfo {author} {\bibfnamefont {L.}~\bibnamefont {Pi}},\ and\
  \bibinfo {author} {\bibfnamefont {Y.}~\bibnamefont {Han}},\ }\href
  {https://doi.org/10.1016/j.jallcom.2024.174478} {\bibfield  {journal}
  {\bibinfo  {journal} {J. Alloys Compd.}\ }\textbf {\bibinfo {volume} {990}},\
  \bibinfo {pages} {174478} (\bibinfo {year} {2024})}\BibitemShut {NoStop}%
\bibitem [{\citenamefont {Rosa}\ \emph {et~al.}(2020)\citenamefont {Rosa},
  \citenamefont {Xu}, \citenamefont {Rahn}, \citenamefont {Souza},
  \citenamefont {Kushwaha}, \citenamefont {Veiga}, \citenamefont {Bombardi},
  \citenamefont {Thomas}, \citenamefont {Janoschek}, \citenamefont {Bauer},
  \citenamefont {Chan}, \citenamefont {Wang}, \citenamefont {Thompson},
  \citenamefont {Harrison}, \citenamefont {Pagliuso}, \citenamefont
  {Bernevig},\ and\ \citenamefont
  {Ronning}}]{Rosa2020Colossalmagnetoresistanceinanonsymmorphic}%
  \BibitemOpen
  \bibfield  {author} {\bibinfo {author} {\bibfnamefont {P.}~\bibnamefont
  {Rosa}}, \bibinfo {author} {\bibfnamefont {Y.}~\bibnamefont {Xu}}, \bibinfo
  {author} {\bibfnamefont {M.}~\bibnamefont {Rahn}}, \bibinfo {author}
  {\bibfnamefont {J.}~\bibnamefont {Souza}}, \bibinfo {author} {\bibfnamefont
  {S.}~\bibnamefont {Kushwaha}}, \bibinfo {author} {\bibfnamefont
  {L.}~\bibnamefont {Veiga}}, \bibinfo {author} {\bibfnamefont
  {A.}~\bibnamefont {Bombardi}}, \bibinfo {author} {\bibfnamefont
  {S.}~\bibnamefont {Thomas}}, \bibinfo {author} {\bibfnamefont
  {M.}~\bibnamefont {Janoschek}}, \bibinfo {author} {\bibfnamefont
  {E.}~\bibnamefont {Bauer}}, \bibinfo {author} {\bibfnamefont
  {M.}~\bibnamefont {Chan}}, \bibinfo {author} {\bibfnamefont {Z.}~\bibnamefont
  {Wang}}, \bibinfo {author} {\bibfnamefont {J.}~\bibnamefont {Thompson}},
  \bibinfo {author} {\bibfnamefont {N.}~\bibnamefont {Harrison}}, \bibinfo
  {author} {\bibfnamefont {P.}~\bibnamefont {Pagliuso}}, \bibinfo {author}
  {\bibfnamefont {A.}~\bibnamefont {Bernevig}},\ and\ \bibinfo {author}
  {\bibfnamefont {F.}~\bibnamefont {Ronning}},\ }\href
  {https://doi.org/10.1038/s41535-020-00256-8} {\bibfield  {journal} {\bibinfo
  {journal} {npj Quantum Mater.}\ }\textbf {\bibinfo {volume} {5}},\ \bibinfo
  {pages} {52} (\bibinfo {year} {2020})}\BibitemShut {NoStop}%
\bibitem [{\citenamefont {Ghosh}\ \emph {et~al.}(2022)\citenamefont {Ghosh},
  \citenamefont {Lane}, \citenamefont {Ronning}, \citenamefont {Bauer},
  \citenamefont {Thompson}, \citenamefont {Zhu}, \citenamefont {Rosa},\ and\
  \citenamefont {Thomas}}]{Ghosh2022Colossalpiezoresistanceinnarrowgap}%
  \BibitemOpen
  \bibfield  {author} {\bibinfo {author} {\bibfnamefont {S.}~\bibnamefont
  {Ghosh}}, \bibinfo {author} {\bibfnamefont {C.}~\bibnamefont {Lane}},
  \bibinfo {author} {\bibfnamefont {F.}~\bibnamefont {Ronning}}, \bibinfo
  {author} {\bibfnamefont {E.~D.}\ \bibnamefont {Bauer}}, \bibinfo {author}
  {\bibfnamefont {J.~D.}\ \bibnamefont {Thompson}}, \bibinfo {author}
  {\bibfnamefont {J.-X.}\ \bibnamefont {Zhu}}, \bibinfo {author} {\bibfnamefont
  {P.~F.~S.}\ \bibnamefont {Rosa}},\ and\ \bibinfo {author} {\bibfnamefont
  {S.~M.}\ \bibnamefont {Thomas}},\ }\href
  {https://doi.org/10.1103/PhysRevB.106.045110} {\bibfield  {journal} {\bibinfo
   {journal} {Phys. Rev. B}\ }\textbf {\bibinfo {volume} {106}},\ \bibinfo
  {pages} {045110} (\bibinfo {year} {2022})}\BibitemShut {NoStop}%
\bibitem [{\citenamefont {Day}\ \emph {et~al.}(2025)\citenamefont {Day},
  \citenamefont {Yamakawa}, \citenamefont {Cairns}, \citenamefont {Singleton},
  \citenamefont {Cao}, \citenamefont {Wu}, \citenamefont {Allen}, \citenamefont
  {Moore},\ and\ \citenamefont
  {Analytis}}]{Day2025Colossalmagnetoresistanceandanisotropic}%
  \BibitemOpen
  \bibfield  {author} {\bibinfo {author} {\bibfnamefont {R.~P.}\ \bibnamefont
  {Day}}, \bibinfo {author} {\bibfnamefont {K.}~\bibnamefont {Yamakawa}},
  \bibinfo {author} {\bibfnamefont {L.~P.}\ \bibnamefont {Cairns}}, \bibinfo
  {author} {\bibfnamefont {J.}~\bibnamefont {Singleton}}, \bibinfo {author}
  {\bibfnamefont {W.~L.}\ \bibnamefont {Cao}}, \bibinfo {author} {\bibfnamefont
  {C.}~\bibnamefont {Wu}}, \bibinfo {author} {\bibfnamefont {M.}~\bibnamefont
  {Allen}}, \bibinfo {author} {\bibfnamefont {J.~E.}\ \bibnamefont {Moore}},\
  and\ \bibinfo {author} {\bibfnamefont {J.~G.}\ \bibnamefont {Analytis}},\
  }\href {https://doi.org/10.1103/PhysRevB.111.054406} {\bibfield  {journal}
  {\bibinfo  {journal} {Phys. Rev. B}\ }\textbf {\bibinfo {volume} {111}},\
  \bibinfo {pages} {054406} (\bibinfo {year} {2025})}\BibitemShut {NoStop}%
\bibitem [{\citenamefont {Balguri}\ \emph {et~al.}(2025)\citenamefont
  {Balguri}, \citenamefont {Mahendru}, \citenamefont {Delgado}, \citenamefont
  {Fruhling}, \citenamefont {Yao}, \citenamefont {Graf}, \citenamefont
  {Rodriguez-Rivera}, \citenamefont {Aczel}, \citenamefont {Hicken},
  \citenamefont {Luetkens}, \citenamefont {Graf}, \citenamefont {Rydh},
  \citenamefont {Gaudet},\ and\ \citenamefont
  {Tafti}}]{Balguri2025Twotypesofcolossalmagnetoresistance}%
  \BibitemOpen
  \bibfield  {author} {\bibinfo {author} {\bibfnamefont {S.}~\bibnamefont
  {Balguri}}, \bibinfo {author} {\bibfnamefont {M.~B.}\ \bibnamefont
  {Mahendru}}, \bibinfo {author} {\bibfnamefont {E.~O.~G.}\ \bibnamefont
  {Delgado}}, \bibinfo {author} {\bibfnamefont {K.}~\bibnamefont {Fruhling}},
  \bibinfo {author} {\bibfnamefont {X.}~\bibnamefont {Yao}}, \bibinfo {author}
  {\bibfnamefont {D.~E.}\ \bibnamefont {Graf}}, \bibinfo {author}
  {\bibfnamefont {J.~A.}\ \bibnamefont {Rodriguez-Rivera}}, \bibinfo {author}
  {\bibfnamefont {A.~A.}\ \bibnamefont {Aczel}}, \bibinfo {author}
  {\bibfnamefont {T.~J.}\ \bibnamefont {Hicken}}, \bibinfo {author}
  {\bibfnamefont {H.}~\bibnamefont {Luetkens}}, \bibinfo {author}
  {\bibfnamefont {M.~J.}\ \bibnamefont {Graf}}, \bibinfo {author}
  {\bibfnamefont {A.}~\bibnamefont {Rydh}}, \bibinfo {author} {\bibfnamefont
  {J.}~\bibnamefont {Gaudet}},\ and\ \bibinfo {author} {\bibfnamefont
  {F.}~\bibnamefont {Tafti}},\ }\href
  {https://doi.org/10.1103/PhysRevB.111.115114} {\bibfield  {journal} {\bibinfo
   {journal} {Phys. Rev. B}\ }\textbf {\bibinfo {volume} {111}},\ \bibinfo
  {pages} {115114} (\bibinfo {year} {2025})}\BibitemShut {NoStop}%
\bibitem [{\citenamefont {Wang}\ \emph {et~al.}(2021)\citenamefont {Wang},
  \citenamefont {Rogers}, \citenamefont {Yao}, \citenamefont {Nichols},
  \citenamefont {Atay}, \citenamefont {Xu}, \citenamefont {Franklin},
  \citenamefont {Sochnikov}, \citenamefont {Ryan}, \citenamefont {Haskel},\
  and\ \citenamefont {Tafti}}]{Wang2021ColossalMagnetoresistancewithoutMixed}%
  \BibitemOpen
  \bibfield  {author} {\bibinfo {author} {\bibfnamefont {Z.-C.}\ \bibnamefont
  {Wang}}, \bibinfo {author} {\bibfnamefont {J.~D.}\ \bibnamefont {Rogers}},
  \bibinfo {author} {\bibfnamefont {X.}~\bibnamefont {Yao}}, \bibinfo {author}
  {\bibfnamefont {R.}~\bibnamefont {Nichols}}, \bibinfo {author} {\bibfnamefont
  {K.}~\bibnamefont {Atay}}, \bibinfo {author} {\bibfnamefont {B.}~\bibnamefont
  {Xu}}, \bibinfo {author} {\bibfnamefont {J.}~\bibnamefont {Franklin}},
  \bibinfo {author} {\bibfnamefont {I.}~\bibnamefont {Sochnikov}}, \bibinfo
  {author} {\bibfnamefont {P.~J.}\ \bibnamefont {Ryan}}, \bibinfo {author}
  {\bibfnamefont {D.}~\bibnamefont {Haskel}},\ and\ \bibinfo {author}
  {\bibfnamefont {F.}~\bibnamefont {Tafti}},\ }\href
  {https://doi.org/https://doi.org/10.1002/adma.202005755} {\bibfield
  {journal} {\bibinfo  {journal} {Adv. Mater.}\ }\textbf {\bibinfo {volume}
  {33}},\ \bibinfo {pages} {2005755} (\bibinfo {year} {2021})}\BibitemShut
  {NoStop}%
\bibitem [{\citenamefont {Usachov}\ \emph {et~al.}(2024)\citenamefont
  {Usachov}, \citenamefont {Krebber}, \citenamefont {Bokai}, \citenamefont
  {Tarasov}, \citenamefont {Kopp}, \citenamefont {Garg}, \citenamefont
  {Virovets}, \citenamefont {M\"uller}, \citenamefont {Mende}, \citenamefont
  {Poelchen}, \citenamefont {Vyalikh}, \citenamefont {Krellner},\ and\
  \citenamefont {Kliemt}}]{Usachov2024Magnetismheatcapacityandelectronic}%
  \BibitemOpen
  \bibfield  {author} {\bibinfo {author} {\bibfnamefont {D.~Y.}\ \bibnamefont
  {Usachov}}, \bibinfo {author} {\bibfnamefont {S.}~\bibnamefont {Krebber}},
  \bibinfo {author} {\bibfnamefont {K.~A.}\ \bibnamefont {Bokai}}, \bibinfo
  {author} {\bibfnamefont {A.~V.}\ \bibnamefont {Tarasov}}, \bibinfo {author}
  {\bibfnamefont {M.}~\bibnamefont {Kopp}}, \bibinfo {author} {\bibfnamefont
  {C.}~\bibnamefont {Garg}}, \bibinfo {author} {\bibfnamefont {A.}~\bibnamefont
  {Virovets}}, \bibinfo {author} {\bibfnamefont {J.}~\bibnamefont {M\"uller}},
  \bibinfo {author} {\bibfnamefont {M.}~\bibnamefont {Mende}}, \bibinfo
  {author} {\bibfnamefont {G.}~\bibnamefont {Poelchen}}, \bibinfo {author}
  {\bibfnamefont {D.~V.}\ \bibnamefont {Vyalikh}}, \bibinfo {author}
  {\bibfnamefont {C.}~\bibnamefont {Krellner}},\ and\ \bibinfo {author}
  {\bibfnamefont {K.}~\bibnamefont {Kliemt}},\ }\href
  {https://doi.org/10.1103/PhysRevB.109.104421} {\bibfield  {journal} {\bibinfo
   {journal} {Phys. Rev. B}\ }\textbf {\bibinfo {volume} {109}},\ \bibinfo
  {pages} {104421} (\bibinfo {year} {2024})}\BibitemShut {NoStop}%
\bibitem [{\citenamefont {Zhang}\ \emph {et~al.}(2023)\citenamefont {Zhang},
  \citenamefont {Du}, \citenamefont {Zheng}, \citenamefont {Luo}, \citenamefont
  {Wu}, \citenamefont {Zheng}, \citenamefont {Cui}, \citenamefont {Sun},
  \citenamefont {Liu}, \citenamefont {Shen}, \citenamefont {Smidman},
  \citenamefont {Song}, \citenamefont {Shi}, \citenamefont {Zhong},
  \citenamefont {Cao}, \citenamefont {Yuan},\ and\ \citenamefont
  {Liu}}]{zhangElectronicBandReconstruction2023a}%
  \BibitemOpen
  \bibfield  {author} {\bibinfo {author} {\bibfnamefont {H.}~\bibnamefont
  {Zhang}}, \bibinfo {author} {\bibfnamefont {F.}~\bibnamefont {Du}}, \bibinfo
  {author} {\bibfnamefont {X.}~\bibnamefont {Zheng}}, \bibinfo {author}
  {\bibfnamefont {S.}~\bibnamefont {Luo}}, \bibinfo {author} {\bibfnamefont
  {Y.}~\bibnamefont {Wu}}, \bibinfo {author} {\bibfnamefont {H.}~\bibnamefont
  {Zheng}}, \bibinfo {author} {\bibfnamefont {S.}~\bibnamefont {Cui}}, \bibinfo
  {author} {\bibfnamefont {Z.}~\bibnamefont {Sun}}, \bibinfo {author}
  {\bibfnamefont {Z.}~\bibnamefont {Liu}}, \bibinfo {author} {\bibfnamefont
  {D.}~\bibnamefont {Shen}}, \bibinfo {author} {\bibfnamefont {M.}~\bibnamefont
  {Smidman}}, \bibinfo {author} {\bibfnamefont {Y.}~\bibnamefont {Song}},
  \bibinfo {author} {\bibfnamefont {M.}~\bibnamefont {Shi}}, \bibinfo {author}
  {\bibfnamefont {Z.}~\bibnamefont {Zhong}}, \bibinfo {author} {\bibfnamefont
  {C.}~\bibnamefont {Cao}}, \bibinfo {author} {\bibfnamefont {H.}~\bibnamefont
  {Yuan}},\ and\ \bibinfo {author} {\bibfnamefont {Y.}~\bibnamefont {Liu}},\
  }\href {https://doi.org/10.1103/PhysRevB.108.L241115} {\bibfield  {journal}
  {\bibinfo  {journal} {Phys. Rev. B}\ }\textbf {\bibinfo {volume} {108}},\
  \bibinfo {pages} {L241115} (\bibinfo {year} {2023})}\BibitemShut {NoStop}%
\bibitem [{\citenamefont {Luo}\ \emph {et~al.}(2023)\citenamefont {Luo},
  \citenamefont {Xu}, \citenamefont {Du}, \citenamefont {Yang}, \citenamefont
  {Chen}, \citenamefont {Cao}, \citenamefont {Song},\ and\ \citenamefont
  {Yuan}}]{Luo2023Colossalmagnetoresistanceandtopologicalphasetransition}%
  \BibitemOpen
  \bibfield  {author} {\bibinfo {author} {\bibfnamefont {S.}~\bibnamefont
  {Luo}}, \bibinfo {author} {\bibfnamefont {Y.}~\bibnamefont {Xu}}, \bibinfo
  {author} {\bibfnamefont {F.}~\bibnamefont {Du}}, \bibinfo {author}
  {\bibfnamefont {L.}~\bibnamefont {Yang}}, \bibinfo {author} {\bibfnamefont
  {Y.}~\bibnamefont {Chen}}, \bibinfo {author} {\bibfnamefont {C.}~\bibnamefont
  {Cao}}, \bibinfo {author} {\bibfnamefont {Y.}~\bibnamefont {Song}},\ and\
  \bibinfo {author} {\bibfnamefont {H.}~\bibnamefont {Yuan}},\ }\href
  {https://doi.org/10.1103/PhysRevB.108.205140} {\bibfield  {journal} {\bibinfo
   {journal} {Phys. Rev. B}\ }\textbf {\bibinfo {volume} {108}},\ \bibinfo
  {pages} {205140} (\bibinfo {year} {2023})}\BibitemShut {NoStop}%
\bibitem [{\citenamefont {Krebber}\ \emph {et~al.}(2023)\citenamefont
  {Krebber}, \citenamefont {Kopp}, \citenamefont {Garg}, \citenamefont
  {Kummer}, \citenamefont {Sichelschmidt}, \citenamefont {Schulz},
  \citenamefont {Poelchen}, \citenamefont {Mende}, \citenamefont {Virovets},
  \citenamefont {Warawa}, \citenamefont {Thomson}, \citenamefont {Tarasov},
  \citenamefont {Usachov}, \citenamefont {Vyalikh}, \citenamefont {Roskos},
  \citenamefont {M\"uller}, \citenamefont {Krellner},\ and\ \citenamefont
  {Kliemt}}]{Krebber2023ColossalmagnetoresistanceinEuZn2P2}%
  \BibitemOpen
  \bibfield  {author} {\bibinfo {author} {\bibfnamefont {S.}~\bibnamefont
  {Krebber}}, \bibinfo {author} {\bibfnamefont {M.}~\bibnamefont {Kopp}},
  \bibinfo {author} {\bibfnamefont {C.}~\bibnamefont {Garg}}, \bibinfo {author}
  {\bibfnamefont {K.}~\bibnamefont {Kummer}}, \bibinfo {author} {\bibfnamefont
  {J.}~\bibnamefont {Sichelschmidt}}, \bibinfo {author} {\bibfnamefont
  {S.}~\bibnamefont {Schulz}}, \bibinfo {author} {\bibfnamefont
  {G.}~\bibnamefont {Poelchen}}, \bibinfo {author} {\bibfnamefont
  {M.}~\bibnamefont {Mende}}, \bibinfo {author} {\bibfnamefont {A.~V.}\
  \bibnamefont {Virovets}}, \bibinfo {author} {\bibfnamefont {K.}~\bibnamefont
  {Warawa}}, \bibinfo {author} {\bibfnamefont {M.~D.}\ \bibnamefont {Thomson}},
  \bibinfo {author} {\bibfnamefont {A.~V.}\ \bibnamefont {Tarasov}}, \bibinfo
  {author} {\bibfnamefont {D.~Y.}\ \bibnamefont {Usachov}}, \bibinfo {author}
  {\bibfnamefont {D.~V.}\ \bibnamefont {Vyalikh}}, \bibinfo {author}
  {\bibfnamefont {H.~G.}\ \bibnamefont {Roskos}}, \bibinfo {author}
  {\bibfnamefont {J.}~\bibnamefont {M\"uller}}, \bibinfo {author}
  {\bibfnamefont {C.}~\bibnamefont {Krellner}},\ and\ \bibinfo {author}
  {\bibfnamefont {K.}~\bibnamefont {Kliemt}},\ }\href
  {https://doi.org/10.1103/PhysRevB.108.045116} {\bibfield  {journal} {\bibinfo
   {journal} {Phys. Rev. B}\ }\textbf {\bibinfo {volume} {108}},\ \bibinfo
  {pages} {045116} (\bibinfo {year} {2023})}\BibitemShut {NoStop}%
\bibitem [{\citenamefont {Krebber}\ \emph {et~al.}(2025)\citenamefont
  {Krebber}, \citenamefont {Sichelschmidt}, \citenamefont {Chailloleau},
  \citenamefont {El~Mard}, \citenamefont {Kopp}, \citenamefont {Baenitz},
  \citenamefont {Kummer}, \citenamefont {Vyalikh}, \citenamefont {M{\"u}ller},
  \citenamefont {Krellner},\ and\ \citenamefont
  {Kliemt}}]{krebberMagneticFrustrationWeak2025a}%
  \BibitemOpen
  \bibfield  {author} {\bibinfo {author} {\bibfnamefont {S.}~\bibnamefont
  {Krebber}}, \bibinfo {author} {\bibfnamefont {J.}~\bibnamefont
  {Sichelschmidt}}, \bibinfo {author} {\bibfnamefont {P.}~\bibnamefont
  {Chailloleau}}, \bibinfo {author} {\bibfnamefont {A.}~\bibnamefont
  {El~Mard}}, \bibinfo {author} {\bibfnamefont {M.}~\bibnamefont {Kopp}},
  \bibinfo {author} {\bibfnamefont {M.}~\bibnamefont {Baenitz}}, \bibinfo
  {author} {\bibfnamefont {K.}~\bibnamefont {Kummer}}, \bibinfo {author}
  {\bibfnamefont {D.~V.}\ \bibnamefont {Vyalikh}}, \bibinfo {author}
  {\bibfnamefont {J.}~\bibnamefont {M{\"u}ller}}, \bibinfo {author}
  {\bibfnamefont {C.}~\bibnamefont {Krellner}},\ and\ \bibinfo {author}
  {\bibfnamefont {K.}~\bibnamefont {Kliemt}},\ }\href
  {https://doi.org/10.1103/PhysRevB.111.174406} {\bibfield  {journal} {\bibinfo
   {journal} {Phys. Rev. B}\ }\textbf {\bibinfo {volume} {111}},\ \bibinfo
  {pages} {174406} (\bibinfo {year} {2025})}\BibitemShut {NoStop}%
\bibitem [{\citenamefont {Chen}\ \emph {et~al.}(2024)\citenamefont {Chen},
  \citenamefont {Dong},\ and\ \citenamefont
  {Wang}}]{chenRecentAdvancesUnderstanding2024}%
  \BibitemOpen
  \bibfield  {author} {\bibinfo {author} {\bibfnamefont {X.}~\bibnamefont
  {Chen}}, \bibinfo {author} {\bibfnamefont {S.}~\bibnamefont {Dong}},\ and\
  \bibinfo {author} {\bibfnamefont {Z.-C.}\ \bibnamefont {Wang}},\ }\href
  {https://doi.org/10.1088/1361-648X/ad882b} {\bibfield  {journal} {\bibinfo
  {journal} {J. Phys. Condens. Matter}\ }\textbf {\bibinfo {volume} {37}},\
  \bibinfo {pages} {033001} (\bibinfo {year} {2024})}\BibitemShut {NoStop}%
\bibitem [{\citenamefont {Kliemt}(2025)}]{kliemt2025}%
  \BibitemOpen
  \bibfield  {author} {\bibinfo {author} {\bibfnamefont {K.}~\bibnamefont
  {Kliemt}},\ }\href {https://doi.org/10.1016/j.pcrysgrow.2025.100667}
  {\bibfield  {journal} {\bibinfo  {journal} {Prog. Cryst. Growth Charact.
  Mater.}\ }\textbf {\bibinfo {volume} {71}},\ \bibinfo {pages} {100667}
  (\bibinfo {year} {2025})}\BibitemShut {NoStop}%
\bibitem [{\citenamefont {Kopp}\ \emph {et~al.}(2026)\citenamefont {Kopp},
  \citenamefont {Garg}, \citenamefont {Krebber}, \citenamefont {Kliemt},
  \citenamefont {Krellner}, \citenamefont {Balguri}, \citenamefont {Mahendru},
  \citenamefont {Tafti}, \citenamefont {Breeze}, \citenamefont {Bentley},
  \citenamefont {Pratt}, \citenamefont {Hicken}, \citenamefont {Luetkens},
  \citenamefont {Krieger}, \citenamefont {Blundell}, \citenamefont {Lancaster},
  \citenamefont {Ale~Crivillero}, \citenamefont {Wirth},\ and\ \citenamefont
  {M{\"u}ller}}]{koppRobustMagneticPolaron2026}%
  \BibitemOpen
  \bibfield  {author} {\bibinfo {author} {\bibfnamefont {M.}~\bibnamefont
  {Kopp}}, \bibinfo {author} {\bibfnamefont {C.}~\bibnamefont {Garg}}, \bibinfo
  {author} {\bibfnamefont {S.}~\bibnamefont {Krebber}}, \bibinfo {author}
  {\bibfnamefont {K.}~\bibnamefont {Kliemt}}, \bibinfo {author} {\bibfnamefont
  {C.}~\bibnamefont {Krellner}}, \bibinfo {author} {\bibfnamefont {S.~R.}\
  \bibnamefont {Balguri}}, \bibinfo {author} {\bibfnamefont {M.}~\bibnamefont
  {Mahendru}}, \bibinfo {author} {\bibfnamefont {F.}~\bibnamefont {Tafti}},
  \bibinfo {author} {\bibfnamefont {T.~L.}\ \bibnamefont {Breeze}}, \bibinfo
  {author} {\bibfnamefont {N.~P.}\ \bibnamefont {Bentley}}, \bibinfo {author}
  {\bibfnamefont {F.~L.}\ \bibnamefont {Pratt}}, \bibinfo {author}
  {\bibfnamefont {T.~J.}\ \bibnamefont {Hicken}}, \bibinfo {author}
  {\bibfnamefont {H.}~\bibnamefont {Luetkens}}, \bibinfo {author}
  {\bibfnamefont {J.~A.}\ \bibnamefont {Krieger}}, \bibinfo {author}
  {\bibfnamefont {S.~J.}\ \bibnamefont {Blundell}}, \bibinfo {author}
  {\bibfnamefont {T.}~\bibnamefont {Lancaster}}, \bibinfo {author}
  {\bibfnamefont {M.~V.}\ \bibnamefont {Ale~Crivillero}}, \bibinfo {author}
  {\bibfnamefont {S.}~\bibnamefont {Wirth}},\ and\ \bibinfo {author}
  {\bibfnamefont {J.}~\bibnamefont {M{\"u}ller}},\ }\href
  {https://doi.org/10.1038/s41535-026-00859-7} {\bibfield  {journal} {\bibinfo
  {journal} {npj Quantum Mater.}\ }\textbf {\bibinfo {volume} {11}},\ \bibinfo
  {pages} {22} (\bibinfo {year} {2026})}\BibitemShut {NoStop}%
\bibitem [{\citenamefont {Cook}\ \emph {et~al.}(2025)\citenamefont {Cook},
  \citenamefont {Peterson}, \citenamefont {Kengle}, \citenamefont {Kennedy},
  \citenamefont {Sheeran}, \citenamefont {Girod}, \citenamefont {Freitas},
  \citenamefont {Greer}, \citenamefont {Abbamonte}, \citenamefont {Pagliuso},
  \citenamefont {Thompson}, \citenamefont {Thomas},\ and\ \citenamefont
  {Rosa}}]{Cook2025MagneticpolaronformationinEuZn2P2}%
  \BibitemOpen
  \bibfield  {author} {\bibinfo {author} {\bibfnamefont {M.~S.}\ \bibnamefont
  {Cook}}, \bibinfo {author} {\bibfnamefont {E.~A.}\ \bibnamefont {Peterson}},
  \bibinfo {author} {\bibfnamefont {C.~S.}\ \bibnamefont {Kengle}}, \bibinfo
  {author} {\bibfnamefont {E.~R.}\ \bibnamefont {Kennedy}}, \bibinfo {author}
  {\bibfnamefont {J.}~\bibnamefont {Sheeran}}, \bibinfo {author} {\bibfnamefont
  {C.}~\bibnamefont {Girod}}, \bibinfo {author} {\bibfnamefont {G.~S.}\
  \bibnamefont {Freitas}}, \bibinfo {author} {\bibfnamefont {S.~M.}\
  \bibnamefont {Greer}}, \bibinfo {author} {\bibfnamefont {P.}~\bibnamefont
  {Abbamonte}}, \bibinfo {author} {\bibfnamefont {P.~G.}\ \bibnamefont
  {Pagliuso}}, \bibinfo {author} {\bibfnamefont {J.~D.}\ \bibnamefont
  {Thompson}}, \bibinfo {author} {\bibfnamefont {S.~M.}\ \bibnamefont
  {Thomas}},\ and\ \bibinfo {author} {\bibfnamefont {P.~F.~S.}\ \bibnamefont
  {Rosa}},\ }\href {https://doi.org/10.1103/fs97-mpcq} {\bibfield  {journal}
  {\bibinfo  {journal} {Phys. Rev. Mater.}\ }\textbf {\bibinfo {volume} {9}},\
  \bibinfo {pages} {104403} (\bibinfo {year} {2025})}\BibitemShut {NoStop}%
\bibitem [{\citenamefont {Li}\ \emph {et~al.}(2024)\citenamefont {Li},
  \citenamefont {Been}, \citenamefont {Balguri}, \citenamefont {Jia},
  \citenamefont {Mahendru}, \citenamefont {Wang}, \citenamefont {Cui},
  \citenamefont {Chen}, \citenamefont {Hashimoto}, \citenamefont {Lu},
  \citenamefont {Moritz}, \citenamefont {Zaanen}, \citenamefont {Tafti},
  \citenamefont {Devereaux},\ and\ \citenamefont
  {Shen}}]{liColossalMagnetoresistanceSpinpolarized2024}%
  \BibitemOpen
  \bibfield  {author} {\bibinfo {author} {\bibfnamefont {Y.-F.}\ \bibnamefont
  {Li}}, \bibinfo {author} {\bibfnamefont {E.~M.}\ \bibnamefont {Been}},
  \bibinfo {author} {\bibfnamefont {S.}~\bibnamefont {Balguri}}, \bibinfo
  {author} {\bibfnamefont {C.-J.}\ \bibnamefont {Jia}}, \bibinfo {author}
  {\bibfnamefont {M.~B.}\ \bibnamefont {Mahendru}}, \bibinfo {author}
  {\bibfnamefont {Z.-C.}\ \bibnamefont {Wang}}, \bibinfo {author}
  {\bibfnamefont {Y.}~\bibnamefont {Cui}}, \bibinfo {author} {\bibfnamefont
  {S.-D.}\ \bibnamefont {Chen}}, \bibinfo {author} {\bibfnamefont
  {M.}~\bibnamefont {Hashimoto}}, \bibinfo {author} {\bibfnamefont {D.-H.}\
  \bibnamefont {Lu}}, \bibinfo {author} {\bibfnamefont {B.}~\bibnamefont
  {Moritz}}, \bibinfo {author} {\bibfnamefont {J.}~\bibnamefont {Zaanen}},
  \bibinfo {author} {\bibfnamefont {F.}~\bibnamefont {Tafti}}, \bibinfo
  {author} {\bibfnamefont {T.~P.}\ \bibnamefont {Devereaux}},\ and\ \bibinfo
  {author} {\bibfnamefont {Z.-X.}\ \bibnamefont {Shen}},\ }\href
  {https://doi.org/10.1073/pnas.2409846121} {\bibfield  {journal} {\bibinfo
  {journal} {PNAS}\ }\textbf {\bibinfo {volume} {121}},\ \bibinfo {pages}
  {e2409846121} (\bibinfo {year} {2024})}\BibitemShut {NoStop}%
\bibitem [{\citenamefont {Ale~Crivillero}\ \emph {et~al.}(2023)\citenamefont
  {Ale~Crivillero}, \citenamefont {R{\"o}{\ss}ler}, \citenamefont {Granovsky},
  \citenamefont {Doerr}, \citenamefont {Cook}, \citenamefont {Rosa},
  \citenamefont {M{\"u}ller},\ and\ \citenamefont
  {Wirth}}]{alecrivilleroMagneticElectronicProperties2023}%
  \BibitemOpen
  \bibfield  {author} {\bibinfo {author} {\bibfnamefont {M.~V.}\ \bibnamefont
  {Ale~Crivillero}}, \bibinfo {author} {\bibfnamefont {S.}~\bibnamefont
  {R{\"o}{\ss}ler}}, \bibinfo {author} {\bibfnamefont {S.}~\bibnamefont
  {Granovsky}}, \bibinfo {author} {\bibfnamefont {M.}~\bibnamefont {Doerr}},
  \bibinfo {author} {\bibfnamefont {M.~S.}\ \bibnamefont {Cook}}, \bibinfo
  {author} {\bibfnamefont {P.~F.~S.}\ \bibnamefont {Rosa}}, \bibinfo {author}
  {\bibfnamefont {J.}~\bibnamefont {M{\"u}ller}},\ and\ \bibinfo {author}
  {\bibfnamefont {S.}~\bibnamefont {Wirth}},\ }\href
  {https://doi.org/10.1038/s41598-023-28711-z} {\bibfield  {journal} {\bibinfo
  {journal} {Sci. Rep.}\ }\textbf {\bibinfo {volume} {13}},\ \bibinfo {pages}
  {1597} (\bibinfo {year} {2023})}\BibitemShut {NoStop}%
\bibitem [{\citenamefont {Sunko}\ \emph {et~al.}(2023)\citenamefont {Sunko},
  \citenamefont {Sun}, \citenamefont {Vranas}, \citenamefont {Homes},
  \citenamefont {Lee}, \citenamefont {Donoway}, \citenamefont {Wang},
  \citenamefont {Balguri}, \citenamefont {Mahendru}, \citenamefont {Ruiz},
  \citenamefont {Gunn}, \citenamefont {Basak}, \citenamefont {{Blanco-Canosa}},
  \citenamefont {Schierle}, \citenamefont {Weschke}, \citenamefont {Tafti},
  \citenamefont {Frano},\ and\ \citenamefont
  {Orenstein}}]{sunkoSpincarrierCouplingInduced2023}%
  \BibitemOpen
  \bibfield  {author} {\bibinfo {author} {\bibfnamefont {V.}~\bibnamefont
  {Sunko}}, \bibinfo {author} {\bibfnamefont {Y.}~\bibnamefont {Sun}}, \bibinfo
  {author} {\bibfnamefont {M.}~\bibnamefont {Vranas}}, \bibinfo {author}
  {\bibfnamefont {C.~C.}\ \bibnamefont {Homes}}, \bibinfo {author}
  {\bibfnamefont {C.}~\bibnamefont {Lee}}, \bibinfo {author} {\bibfnamefont
  {E.}~\bibnamefont {Donoway}}, \bibinfo {author} {\bibfnamefont {Z.-C.}\
  \bibnamefont {Wang}}, \bibinfo {author} {\bibfnamefont {S.}~\bibnamefont
  {Balguri}}, \bibinfo {author} {\bibfnamefont {M.~B.}\ \bibnamefont
  {Mahendru}}, \bibinfo {author} {\bibfnamefont {A.}~\bibnamefont {Ruiz}},
  \bibinfo {author} {\bibfnamefont {B.}~\bibnamefont {Gunn}}, \bibinfo {author}
  {\bibfnamefont {R.}~\bibnamefont {Basak}}, \bibinfo {author} {\bibfnamefont
  {S.}~\bibnamefont {{Blanco-Canosa}}}, \bibinfo {author} {\bibfnamefont
  {E.}~\bibnamefont {Schierle}}, \bibinfo {author} {\bibfnamefont
  {E.}~\bibnamefont {Weschke}}, \bibinfo {author} {\bibfnamefont
  {F.}~\bibnamefont {Tafti}}, \bibinfo {author} {\bibfnamefont
  {A.}~\bibnamefont {Frano}},\ and\ \bibinfo {author} {\bibfnamefont
  {J.}~\bibnamefont {Orenstein}},\ }\href
  {https://doi.org/10.1103/PhysRevB.107.144404} {\bibfield  {journal} {\bibinfo
   {journal} {Phys. Rev. B}\ }\textbf {\bibinfo {volume} {107}},\ \bibinfo
  {pages} {144404} (\bibinfo {year} {2023})}\BibitemShut {NoStop}%
\bibitem [{\citenamefont {Sar}\ \emph {et~al.}(2026)\citenamefont {Sar},
  \citenamefont {Gong}, \citenamefont {Romanova}, \citenamefont
  {Khizanishvili}, \citenamefont {Park}, \citenamefont {Lee}, \citenamefont
  {Kaczorowski},\ and\ \citenamefont
  {Aynajian}}]{sarTopologicallyTrivialSemiconducting2026}%
  \BibitemOpen
  \bibfield  {author} {\bibinfo {author} {\bibfnamefont {D.}~\bibnamefont
  {Sar}}, \bibinfo {author} {\bibfnamefont {M.}~\bibnamefont {Gong}}, \bibinfo
  {author} {\bibfnamefont {T.}~\bibnamefont {Romanova}}, \bibinfo {author}
  {\bibfnamefont {L.}~\bibnamefont {Khizanishvili}}, \bibinfo {author}
  {\bibfnamefont {H.}~\bibnamefont {Park}}, \bibinfo {author} {\bibfnamefont
  {W.-C.}\ \bibnamefont {Lee}}, \bibinfo {author} {\bibfnamefont
  {D.}~\bibnamefont {Kaczorowski}},\ and\ \bibinfo {author} {\bibfnamefont
  {P.}~\bibnamefont {Aynajian}},\ }\href
  {https://doi.org/10.1088/1367-2630/ae33e3} {\bibfield  {journal} {\bibinfo
  {journal} {New J. Phys.}\ }\textbf {\bibinfo {volume} {28}},\ \bibinfo
  {pages} {013502} (\bibinfo {year} {2026})}\BibitemShut {NoStop}%
\bibitem [{\citenamefont {Wang}\ \emph {et~al.}(2011)\citenamefont {Wang},
  \citenamefont {Yang}, \citenamefont {Pan}, \citenamefont {Xia}, \citenamefont
  {Tao}, \citenamefont {He}, \citenamefont {Darone},\ and\ \citenamefont
  {Bobev}}]{Wang2011SynthesisCrystalandElectronicStructures}%
  \BibitemOpen
  \bibfield  {author} {\bibinfo {author} {\bibfnamefont {J.}~\bibnamefont
  {Wang}}, \bibinfo {author} {\bibfnamefont {M.}~\bibnamefont {Yang}}, \bibinfo
  {author} {\bibfnamefont {M.-Y.}\ \bibnamefont {Pan}}, \bibinfo {author}
  {\bibfnamefont {S.-Q.}\ \bibnamefont {Xia}}, \bibinfo {author} {\bibfnamefont
  {X.-T.}\ \bibnamefont {Tao}}, \bibinfo {author} {\bibfnamefont
  {H.}~\bibnamefont {He}}, \bibinfo {author} {\bibfnamefont {G.}~\bibnamefont
  {Darone}},\ and\ \bibinfo {author} {\bibfnamefont {S.}~\bibnamefont
  {Bobev}},\ }\href {https://doi.org/10.1021/ic200286t} {\bibfield  {journal}
  {\bibinfo  {journal} {Inorg. Chem.}\ }\textbf {\bibinfo {volume} {50}},\
  \bibinfo {pages} {8020} (\bibinfo {year} {2011})}\BibitemShut {NoStop}%
\bibitem [{\citenamefont {Dolomanov}\ \emph {et~al.}(2009)\citenamefont
  {Dolomanov}, \citenamefont {Bourhis}, \citenamefont {Gildea}, \citenamefont
  {Howard},\ and\ \citenamefont {Puschmann}}]{26}%
  \BibitemOpen
  \bibfield  {author} {\bibinfo {author} {\bibfnamefont {O.~V.}\ \bibnamefont
  {Dolomanov}}, \bibinfo {author} {\bibfnamefont {L.~J.}\ \bibnamefont
  {Bourhis}}, \bibinfo {author} {\bibfnamefont {R.~J.}\ \bibnamefont {Gildea}},
  \bibinfo {author} {\bibfnamefont {J.~A.}\ \bibnamefont {Howard}},\ and\
  \bibinfo {author} {\bibfnamefont {H.}~\bibnamefont {Puschmann}},\ }\href
  {https://doi.org/10.1107/S0021889808042726} {\bibfield  {journal} {\bibinfo
  {journal} {J. Appl. Crystallogr.}\ }\textbf {\bibinfo {volume} {42}},\
  \bibinfo {pages} {339} (\bibinfo {year} {2009})}\BibitemShut {NoStop}%
\bibitem [{\citenamefont {H{\"u}bschle}\ \emph {et~al.}(2011)\citenamefont
  {H{\"u}bschle}, \citenamefont {Sheldrick},\ and\ \citenamefont
  {Dittrich}}]{27}%
  \BibitemOpen
  \bibfield  {author} {\bibinfo {author} {\bibfnamefont {C.~B.}\ \bibnamefont
  {H{\"u}bschle}}, \bibinfo {author} {\bibfnamefont {G.~M.}\ \bibnamefont
  {Sheldrick}},\ and\ \bibinfo {author} {\bibfnamefont {B.}~\bibnamefont
  {Dittrich}},\ }\href {https://doi.org/10.1107/S0021889811043202} {\bibfield
  {journal} {\bibinfo  {journal} {J. Appl. Crystallogr.}\ }\textbf {\bibinfo
  {volume} {44}},\ \bibinfo {pages} {1281} (\bibinfo {year}
  {2011})}\BibitemShut {NoStop}%
\bibitem [{\citenamefont {Momma}\ and\ \citenamefont {Izumi}(2011)}]{28}%
  \BibitemOpen
  \bibfield  {author} {\bibinfo {author} {\bibfnamefont {K.}~\bibnamefont
  {Momma}}\ and\ \bibinfo {author} {\bibfnamefont {F.}~\bibnamefont {Izumi}},\
  }\href {https://doi.org/10.1107/S0021889811043202} {\bibfield  {journal}
  {\bibinfo  {journal} {J. Appl. Crystallogr.}\ }\textbf {\bibinfo {volume}
  {44}},\ \bibinfo {pages} {1272} (\bibinfo {year} {2011})}\BibitemShut
  {NoStop}%
\bibitem [{SM()}]{SM}%
  \BibitemOpen
  \href {https://journals.aps.org} {}\bibinfo {note} {See Supplemental Material
  at http://link.aps.org/supplemental/xxxxxx for supplementary
  crystallographic, magnetic, heat capacity, and magnetotransport data for
  Eu$_2$CdAs$_2$.}\BibitemShut {Stop}%
\bibitem [{\citenamefont {Fender}\ \emph {et~al.}(2021)\citenamefont {Fender},
  \citenamefont {Thomas}, \citenamefont {Ronning}, \citenamefont {Bauer},
  \citenamefont {Thompson},\ and\ \citenamefont
  {Rosa}}]{Fender2021Narrowgapsemiconductingbehavior}%
  \BibitemOpen
  \bibfield  {author} {\bibinfo {author} {\bibfnamefont {S.~S.}\ \bibnamefont
  {Fender}}, \bibinfo {author} {\bibfnamefont {S.~M.}\ \bibnamefont {Thomas}},
  \bibinfo {author} {\bibfnamefont {F.}~\bibnamefont {Ronning}}, \bibinfo
  {author} {\bibfnamefont {E.~D.}\ \bibnamefont {Bauer}}, \bibinfo {author}
  {\bibfnamefont {J.~D.}\ \bibnamefont {Thompson}},\ and\ \bibinfo {author}
  {\bibfnamefont {P.~F.~S.}\ \bibnamefont {Rosa}},\ }\href
  {https://doi.org/10.1103/PhysRevMaterials.5.074603} {\bibfield  {journal}
  {\bibinfo  {journal} {Phys. Rev. Mater.}\ }\textbf {\bibinfo {volume} {5}},\
  \bibinfo {pages} {074603} (\bibinfo {year} {2021})}\BibitemShut {NoStop}%
\bibitem [{\citenamefont {{Santos-Cottin}}\ \emph {et~al.}(2023)\citenamefont
  {{Santos-Cottin}}, \citenamefont {Mohelsk{\'y}}, \citenamefont {Wyzula},
  \citenamefont {Le~Mardel{\'e}}, \citenamefont {Kapon}, \citenamefont
  {Nasrallah}, \citenamefont {Bari{\v s}i{\'c}}, \citenamefont {{\v
  Z}ivkovi{\'c}}, \citenamefont {Soh}, \citenamefont {Guo}, \citenamefont
  {Rigaux}, \citenamefont {Puppin}, \citenamefont {Dil}, \citenamefont {Gudac},
  \citenamefont {Rukelj}, \citenamefont {Novak}, \citenamefont {Kuzmenko},
  \citenamefont {Homes}, \citenamefont {Dietl}, \citenamefont {Orlita},\ and\
  \citenamefont {Akrap}}]{santos-cottin2023}%
  \BibitemOpen
  \bibfield  {author} {\bibinfo {author} {\bibfnamefont {D.}~\bibnamefont
  {{Santos-Cottin}}}, \bibinfo {author} {\bibfnamefont {I.}~\bibnamefont
  {Mohelsk{\'y}}}, \bibinfo {author} {\bibfnamefont {J.}~\bibnamefont
  {Wyzula}}, \bibinfo {author} {\bibfnamefont {F.}~\bibnamefont
  {Le~Mardel{\'e}}}, \bibinfo {author} {\bibfnamefont {I.}~\bibnamefont
  {Kapon}}, \bibinfo {author} {\bibfnamefont {S.}~\bibnamefont {Nasrallah}},
  \bibinfo {author} {\bibfnamefont {N.}~\bibnamefont {Bari{\v s}i{\'c}}},
  \bibinfo {author} {\bibfnamefont {I.}~\bibnamefont {{\v Z}ivkovi{\'c}}},
  \bibinfo {author} {\bibfnamefont {J.~R.}\ \bibnamefont {Soh}}, \bibinfo
  {author} {\bibfnamefont {F.}~\bibnamefont {Guo}}, \bibinfo {author}
  {\bibfnamefont {K.}~\bibnamefont {Rigaux}}, \bibinfo {author} {\bibfnamefont
  {M.}~\bibnamefont {Puppin}}, \bibinfo {author} {\bibfnamefont {J.~H.}\
  \bibnamefont {Dil}}, \bibinfo {author} {\bibfnamefont {B.}~\bibnamefont
  {Gudac}}, \bibinfo {author} {\bibfnamefont {Z.}~\bibnamefont {Rukelj}},
  \bibinfo {author} {\bibfnamefont {M.}~\bibnamefont {Novak}}, \bibinfo
  {author} {\bibfnamefont {A.~B.}\ \bibnamefont {Kuzmenko}}, \bibinfo {author}
  {\bibfnamefont {C.~C.}\ \bibnamefont {Homes}}, \bibinfo {author}
  {\bibfnamefont {T.}~\bibnamefont {Dietl}}, \bibinfo {author} {\bibfnamefont
  {M.}~\bibnamefont {Orlita}},\ and\ \bibinfo {author} {\bibfnamefont
  {A.}~\bibnamefont {Akrap}},\ }\href
  {https://doi.org/10.1103/PhysRevLett.131.186704} {\bibfield  {journal}
  {\bibinfo  {journal} {Phys. Rev. Lett.}\ }\textbf {\bibinfo {volume} {131}},\
  \bibinfo {pages} {186704} (\bibinfo {year} {2023})}\BibitemShut {NoStop}%
\bibitem [{\citenamefont {Yi}\ \emph {et~al.}(2017)\citenamefont {Yi},
  \citenamefont {Yang}, \citenamefont {Yang}, \citenamefont {Wang},
  \citenamefont {Matsushita}, \citenamefont {Miao}, \citenamefont {Jiao},
  \citenamefont {Cheng}, \citenamefont {Li}, \citenamefont {Yamaura},
  \citenamefont {Shi},\ and\ \citenamefont {Luo}}]{yi2017}%
  \BibitemOpen
  \bibfield  {author} {\bibinfo {author} {\bibfnamefont {C.}~\bibnamefont
  {Yi}}, \bibinfo {author} {\bibfnamefont {S.}~\bibnamefont {Yang}}, \bibinfo
  {author} {\bibfnamefont {M.}~\bibnamefont {Yang}}, \bibinfo {author}
  {\bibfnamefont {L.}~\bibnamefont {Wang}}, \bibinfo {author} {\bibfnamefont
  {Y.}~\bibnamefont {Matsushita}}, \bibinfo {author} {\bibfnamefont
  {S.}~\bibnamefont {Miao}}, \bibinfo {author} {\bibfnamefont {Y.}~\bibnamefont
  {Jiao}}, \bibinfo {author} {\bibfnamefont {J.}~\bibnamefont {Cheng}},
  \bibinfo {author} {\bibfnamefont {Y.}~\bibnamefont {Li}}, \bibinfo {author}
  {\bibfnamefont {K.}~\bibnamefont {Yamaura}}, \bibinfo {author} {\bibfnamefont
  {Y.}~\bibnamefont {Shi}},\ and\ \bibinfo {author} {\bibfnamefont
  {J.}~\bibnamefont {Luo}},\ }\href
  {https://doi.org/10.1103/PhysRevB.96.205103} {\bibfield  {journal} {\bibinfo
  {journal} {Phys. Rev. B}\ }\textbf {\bibinfo {volume} {96}},\ \bibinfo
  {pages} {205103} (\bibinfo {year} {2017})}\BibitemShut {NoStop}%
\bibitem [{\citenamefont {Iwamoto}\ \emph {et~al.}(2022)\citenamefont
  {Iwamoto}, \citenamefont {Sekino}, \citenamefont {Ito}, \citenamefont
  {Katayama}, \citenamefont {Ueno},\ and\ \citenamefont
  {Katsufuji}}]{Iwamoto2022LargeNegativeMagnetoresistance}%
  \BibitemOpen
  \bibfield  {author} {\bibinfo {author} {\bibfnamefont {K.}~\bibnamefont
  {Iwamoto}}, \bibinfo {author} {\bibfnamefont {W.}~\bibnamefont {Sekino}},
  \bibinfo {author} {\bibfnamefont {S.}~\bibnamefont {Ito}}, \bibinfo {author}
  {\bibfnamefont {Y.}~\bibnamefont {Katayama}}, \bibinfo {author}
  {\bibfnamefont {K.}~\bibnamefont {Ueno}},\ and\ \bibinfo {author}
  {\bibfnamefont {T.}~\bibnamefont {Katsufuji}},\ }\href
  {https://doi.org/10.7566/JPSJ.91.033702} {\bibfield  {journal} {\bibinfo
  {journal} {J. Phys. Soc. Jpn.}\ }\textbf {\bibinfo {volume} {91}},\ \bibinfo
  {pages} {033702} (\bibinfo {year} {2022})}\BibitemShut {NoStop}%
\bibitem [{\citenamefont {Glushkov}\ \emph {et~al.}(2010)\citenamefont
  {Glushkov}, \citenamefont {Anisimov}, \citenamefont {Bogach}, \citenamefont
  {Demishev}, \citenamefont {Dukhnenko}, \citenamefont {Kuznetsov},
  \citenamefont {Levchenko}, \citenamefont {Samarin}, \citenamefont {Filippov},
  \citenamefont {Churkin}, \citenamefont {Shitsevalova},\ and\ \citenamefont
  {Sluchanko}}]{glushkov2010Enhancementofthecolossalmagnetoresistance}%
  \BibitemOpen
  \bibfield  {author} {\bibinfo {author} {\bibfnamefont {V.~V.}\ \bibnamefont
  {Glushkov}}, \bibinfo {author} {\bibfnamefont {M.~A.}\ \bibnamefont
  {Anisimov}}, \bibinfo {author} {\bibfnamefont {A.~V.}\ \bibnamefont
  {Bogach}}, \bibinfo {author} {\bibfnamefont {S.~V.}\ \bibnamefont
  {Demishev}}, \bibinfo {author} {\bibfnamefont {A.~V.}\ \bibnamefont
  {Dukhnenko}}, \bibinfo {author} {\bibfnamefont {A.~V.}\ \bibnamefont
  {Kuznetsov}}, \bibinfo {author} {\bibfnamefont {A.~V.}\ \bibnamefont
  {Levchenko}}, \bibinfo {author} {\bibfnamefont {N.~A.}\ \bibnamefont
  {Samarin}}, \bibinfo {author} {\bibfnamefont {V.~B.}\ \bibnamefont
  {Filippov}}, \bibinfo {author} {\bibfnamefont {O.~A.}\ \bibnamefont
  {Churkin}}, \bibinfo {author} {\bibfnamefont {N.~Y.}\ \bibnamefont
  {Shitsevalova}},\ and\ \bibinfo {author} {\bibfnamefont {N.~E.}\ \bibnamefont
  {Sluchanko}},\ }\href {https://doi.org/10.1134/S1063776110080157} {\bibfield
  {journal} {\bibinfo  {journal} {J. Exp. Theor. Phys.}\ }\textbf {\bibinfo
  {volume} {111}},\ \bibinfo {pages} {246} (\bibinfo {year}
  {2010})}\BibitemShut {NoStop}%
\bibitem [{\citenamefont {Devlin}\ \emph {et~al.}(2018)\citenamefont {Devlin},
  \citenamefont {Kazem}, \citenamefont {Zaikina}, \citenamefont {Cooley},
  \citenamefont {Badger}, \citenamefont {Fettinger}, \citenamefont {Taufour},\
  and\ \citenamefont {Kauzlarich}}]{Devlin2018Eu11Zn4Sn2As12}%
  \BibitemOpen
  \bibfield  {author} {\bibinfo {author} {\bibfnamefont {K.~P.}\ \bibnamefont
  {Devlin}}, \bibinfo {author} {\bibfnamefont {N.}~\bibnamefont {Kazem}},
  \bibinfo {author} {\bibfnamefont {J.~V.}\ \bibnamefont {Zaikina}}, \bibinfo
  {author} {\bibfnamefont {J.~A.}\ \bibnamefont {Cooley}}, \bibinfo {author}
  {\bibfnamefont {J.~R.}\ \bibnamefont {Badger}}, \bibinfo {author}
  {\bibfnamefont {J.~C.}\ \bibnamefont {Fettinger}}, \bibinfo {author}
  {\bibfnamefont {V.}~\bibnamefont {Taufour}},\ and\ \bibinfo {author}
  {\bibfnamefont {S.~M.}\ \bibnamefont {Kauzlarich}},\ }\href
  {https://doi.org/10.1021/acs.chemmater.8b02749} {\bibfield  {journal}
  {\bibinfo  {journal} {Chem. Mater.}\ }\textbf {\bibinfo {volume} {30}},\
  \bibinfo {pages} {7067} (\bibinfo {year} {2018})}\BibitemShut {NoStop}%
\bibitem [{\citenamefont {Shi}\ \emph {et~al.}(2024)\citenamefont {Shi},
  \citenamefont {Liu}, \citenamefont {Burnett}, \citenamefont {Lee},
  \citenamefont {Hu}, \citenamefont {Jiang}, \citenamefont {Cai}, \citenamefont
  {Xu}, \citenamefont {Li}, \citenamefont {Chen},\ and\ \citenamefont
  {Chu}}]{Shi2024AbsenceofWeylnodesinEuCd2As2}%
  \BibitemOpen
  \bibfield  {author} {\bibinfo {author} {\bibfnamefont {Y.}~\bibnamefont
  {Shi}}, \bibinfo {author} {\bibfnamefont {Z.}~\bibnamefont {Liu}}, \bibinfo
  {author} {\bibfnamefont {L.~A.}\ \bibnamefont {Burnett}}, \bibinfo {author}
  {\bibfnamefont {S.}~\bibnamefont {Lee}}, \bibinfo {author} {\bibfnamefont
  {C.}~\bibnamefont {Hu}}, \bibinfo {author} {\bibfnamefont {Q.}~\bibnamefont
  {Jiang}}, \bibinfo {author} {\bibfnamefont {J.}~\bibnamefont {Cai}}, \bibinfo
  {author} {\bibfnamefont {X.}~\bibnamefont {Xu}}, \bibinfo {author}
  {\bibfnamefont {M.}~\bibnamefont {Li}}, \bibinfo {author} {\bibfnamefont
  {C.-C.}\ \bibnamefont {Chen}},\ and\ \bibinfo {author} {\bibfnamefont
  {J.-H.}\ \bibnamefont {Chu}},\ }\href
  {https://doi.org/10.1103/PhysRevB.109.125202} {\bibfield  {journal} {\bibinfo
   {journal} {Phys. Rev. B}\ }\textbf {\bibinfo {volume} {109}},\ \bibinfo
  {pages} {125202} (\bibinfo {year} {2024})}\BibitemShut {NoStop}%
\bibitem [{\citenamefont {Roychowdhury}\ \emph {et~al.}(2023)\citenamefont
  {Roychowdhury}, \citenamefont {Yao}, \citenamefont {Samanta}, \citenamefont
  {Bae}, \citenamefont {Chen}, \citenamefont {Ju}, \citenamefont {Raghavan},
  \citenamefont {Kumar}, \citenamefont {Constantinou}, \citenamefont {Guin},
  \citenamefont {Plumb}, \citenamefont {Romanelli}, \citenamefont {Borrmann},
  \citenamefont {Vergniory}, \citenamefont {Strocov}, \citenamefont {Madhavan},
  \citenamefont {Shekhar},\ and\ \citenamefont
  {Felser}}]{Roychowdhury2023AnomalousHallConductivityandNernstEffect}%
  \BibitemOpen
  \bibfield  {author} {\bibinfo {author} {\bibfnamefont {S.}~\bibnamefont
  {Roychowdhury}}, \bibinfo {author} {\bibfnamefont {M.}~\bibnamefont {Yao}},
  \bibinfo {author} {\bibfnamefont {K.}~\bibnamefont {Samanta}}, \bibinfo
  {author} {\bibfnamefont {S.}~\bibnamefont {Bae}}, \bibinfo {author}
  {\bibfnamefont {D.}~\bibnamefont {Chen}}, \bibinfo {author} {\bibfnamefont
  {S.}~\bibnamefont {Ju}}, \bibinfo {author} {\bibfnamefont {A.}~\bibnamefont
  {Raghavan}}, \bibinfo {author} {\bibfnamefont {N.}~\bibnamefont {Kumar}},
  \bibinfo {author} {\bibfnamefont {P.}~\bibnamefont {Constantinou}}, \bibinfo
  {author} {\bibfnamefont {S.~N.}\ \bibnamefont {Guin}}, \bibinfo {author}
  {\bibfnamefont {N.~C.}\ \bibnamefont {Plumb}}, \bibinfo {author}
  {\bibfnamefont {M.}~\bibnamefont {Romanelli}}, \bibinfo {author}
  {\bibfnamefont {H.}~\bibnamefont {Borrmann}}, \bibinfo {author}
  {\bibfnamefont {M.~G.}\ \bibnamefont {Vergniory}}, \bibinfo {author}
  {\bibfnamefont {V.~N.}\ \bibnamefont {Strocov}}, \bibinfo {author}
  {\bibfnamefont {V.}~\bibnamefont {Madhavan}}, \bibinfo {author}
  {\bibfnamefont {C.}~\bibnamefont {Shekhar}},\ and\ \bibinfo {author}
  {\bibfnamefont {C.}~\bibnamefont {Felser}},\ }\href
  {https://doi.org/https://doi.org/10.1002/advs.202207121} {\bibfield
  {journal} {\bibinfo  {journal} {Adv. Sci.}\ }\textbf {\bibinfo {volume}
  {10}},\ \bibinfo {pages} {2207121} (\bibinfo {year} {2023})}\BibitemShut
  {NoStop}%
\bibitem [{\citenamefont {Dong}\ \emph {et~al.}(2025)\citenamefont {Dong},
  \citenamefont {Yang}, \citenamefont {Liu}, \citenamefont {Wang},
  \citenamefont {Wen}, \citenamefont {Liu}, \citenamefont {Shi}, \citenamefont
  {Tian}, \citenamefont {Sun}, \citenamefont {Uwatoko}, \citenamefont {Wu},
  \citenamefont {Chen}, \citenamefont {Wang},\ and\ \citenamefont
  {Cheng}}]{Dong2025Simultaneouscolossalmagnetoresistanceandangular}%
  \BibitemOpen
  \bibfield  {author} {\bibinfo {author} {\bibfnamefont {Q.}~\bibnamefont
  {Dong}}, \bibinfo {author} {\bibfnamefont {P.}~\bibnamefont {Yang}}, \bibinfo
  {author} {\bibfnamefont {Z.}~\bibnamefont {Liu}}, \bibinfo {author}
  {\bibfnamefont {Y.}~\bibnamefont {Wang}}, \bibinfo {author} {\bibfnamefont
  {H.}~\bibnamefont {Wen}}, \bibinfo {author} {\bibfnamefont {Z.}~\bibnamefont
  {Liu}}, \bibinfo {author} {\bibfnamefont {T.}~\bibnamefont {Shi}}, \bibinfo
  {author} {\bibfnamefont {Z.}~\bibnamefont {Tian}}, \bibinfo {author}
  {\bibfnamefont {J.}~\bibnamefont {Sun}}, \bibinfo {author} {\bibfnamefont
  {Y.}~\bibnamefont {Uwatoko}}, \bibinfo {author} {\bibfnamefont
  {Q.}~\bibnamefont {Wu}}, \bibinfo {author} {\bibfnamefont {G.}~\bibnamefont
  {Chen}}, \bibinfo {author} {\bibfnamefont {B.}~\bibnamefont {Wang}},\ and\
  \bibinfo {author} {\bibfnamefont {J.}~\bibnamefont {Cheng}},\ }\href
  {https://doi.org/10.1103/p2c5-r163} {\bibfield  {journal} {\bibinfo
  {journal} {Phys. Rev. B}\ }\textbf {\bibinfo {volume} {112}},\ \bibinfo
  {pages} {L140405} (\bibinfo {year} {2025})}\BibitemShut {NoStop}%
\bibitem [{\citenamefont {Yin}\ \emph {et~al.}(2020)\citenamefont {Yin},
  \citenamefont {Wu}, \citenamefont {Li}, \citenamefont {Yu}, \citenamefont
  {Sun}, \citenamefont {Shen}, \citenamefont {Frandsen}, \citenamefont {Yao},\
  and\ \citenamefont
  {Wang}}]{Yin2020Largenegativemagnetoresistanceintheantiferromagnetic}%
  \BibitemOpen
  \bibfield  {author} {\bibinfo {author} {\bibfnamefont {J.}~\bibnamefont
  {Yin}}, \bibinfo {author} {\bibfnamefont {C.}~\bibnamefont {Wu}}, \bibinfo
  {author} {\bibfnamefont {L.}~\bibnamefont {Li}}, \bibinfo {author}
  {\bibfnamefont {J.}~\bibnamefont {Yu}}, \bibinfo {author} {\bibfnamefont
  {H.}~\bibnamefont {Sun}}, \bibinfo {author} {\bibfnamefont {B.}~\bibnamefont
  {Shen}}, \bibinfo {author} {\bibfnamefont {B.~A.}\ \bibnamefont {Frandsen}},
  \bibinfo {author} {\bibfnamefont {D.-X.}\ \bibnamefont {Yao}},\ and\ \bibinfo
  {author} {\bibfnamefont {M.}~\bibnamefont {Wang}},\ }\href
  {https://doi.org/10.1103/PhysRevMaterials.4.013405} {\bibfield  {journal}
  {\bibinfo  {journal} {Phys. Rev. Mater.}\ }\textbf {\bibinfo {volume} {4}},\
  \bibinfo {pages} {013405} (\bibinfo {year} {2020})}\BibitemShut {NoStop}%
\bibitem [{\citenamefont {Chen}\ \emph {et~al.}(2018)\citenamefont {Chen},
  \citenamefont {Ren}, \citenamefont {Rubi}, \citenamefont {Mahendiran},\ and\
  \citenamefont {Wang}}]{chen2018Strongelectronphononinteractionandcolossal}%
  \BibitemOpen
  \bibfield  {author} {\bibinfo {author} {\bibfnamefont {R.}~\bibnamefont
  {Chen}}, \bibinfo {author} {\bibfnamefont {J.-C.}\ \bibnamefont {Ren}},
  \bibinfo {author} {\bibfnamefont {K.}~\bibnamefont {Rubi}}, \bibinfo {author}
  {\bibfnamefont {R.}~\bibnamefont {Mahendiran}},\ and\ \bibinfo {author}
  {\bibfnamefont {J.-S.}\ \bibnamefont {Wang}},\ }\href
  {https://doi.org/10.1140/epjb/e2018-80536-8} {\bibfield  {journal} {\bibinfo
  {journal} {Eur. Phys. J. B}\ }\textbf {\bibinfo {volume} {91}},\ \bibinfo
  {pages} {36} (\bibinfo {year} {2018})}\BibitemShut {NoStop}%
\bibitem [{\citenamefont {Kim}\ \emph {et~al.}(2002)\citenamefont {Kim},
  \citenamefont {Olmstead}, \citenamefont {Klavins}, \citenamefont {Webb},\
  and\ \citenamefont {Kauzlarich}}]{Kim2002StructureMagnetismandColossal}%
  \BibitemOpen
  \bibfield  {author} {\bibinfo {author} {\bibfnamefont {H.}~\bibnamefont
  {Kim}}, \bibinfo {author} {\bibfnamefont {M.~M.}\ \bibnamefont {Olmstead}},
  \bibinfo {author} {\bibfnamefont {P.}~\bibnamefont {Klavins}}, \bibinfo
  {author} {\bibfnamefont {D.~J.}\ \bibnamefont {Webb}},\ and\ \bibinfo
  {author} {\bibfnamefont {S.~M.}\ \bibnamefont {Kauzlarich}},\ }\href
  {https://doi.org/10.1021/cm020065w} {\bibfield  {journal} {\bibinfo
  {journal} {Chem. Mater.}\ }\textbf {\bibinfo {volume} {14}},\ \bibinfo
  {pages} {3382} (\bibinfo {year} {2002})}\BibitemShut {NoStop}%
\bibitem [{\citenamefont {Rubi}\ and\ \citenamefont
  {Mahendiran}(2017)}]{Rubi2017Largemagnetoresistanceoverawide}%
  \BibitemOpen
  \bibfield  {author} {\bibinfo {author} {\bibfnamefont {K.}~\bibnamefont
  {Rubi}}\ and\ \bibinfo {author} {\bibfnamefont {R.}~\bibnamefont
  {Mahendiran}},\ }\href {https://doi.org/10.1209/0295-5075/118/57008}
  {\bibfield  {journal} {\bibinfo  {journal} {Europhys. Lett.}\ }\textbf
  {\bibinfo {volume} {118}},\ \bibinfo {pages} {57008} (\bibinfo {year}
  {2017})}\BibitemShut {NoStop}%
\bibitem [{\citenamefont {Glushkov}\ \emph {et~al.}(2007)\citenamefont
  {Glushkov}, \citenamefont {Bogach}, \citenamefont {Gon’kov}, \citenamefont
  {Demishev}, \citenamefont {Ivanov}, \citenamefont {Kuznetsov}, \citenamefont
  {Samarin}, \citenamefont {Shitsevalova}, \citenamefont {Flachbart},\ and\
  \citenamefont {Sluchanko}}]{glushkov2007Magneticandtransportproperties}%
  \BibitemOpen
  \bibfield  {author} {\bibinfo {author} {\bibfnamefont {V.~V.}\ \bibnamefont
  {Glushkov}}, \bibinfo {author} {\bibfnamefont {A.~V.}\ \bibnamefont
  {Bogach}}, \bibinfo {author} {\bibfnamefont {K.~V.}\ \bibnamefont
  {Gon’kov}}, \bibinfo {author} {\bibfnamefont {S.~V.}\ \bibnamefont
  {Demishev}}, \bibinfo {author} {\bibfnamefont {V.~Y.}\ \bibnamefont
  {Ivanov}}, \bibinfo {author} {\bibfnamefont {A.~V.}\ \bibnamefont
  {Kuznetsov}}, \bibinfo {author} {\bibfnamefont {N.~A.}\ \bibnamefont
  {Samarin}}, \bibinfo {author} {\bibfnamefont {N.~Y.}\ \bibnamefont
  {Shitsevalova}}, \bibinfo {author} {\bibfnamefont {K.}~\bibnamefont
  {Flachbart}},\ and\ \bibinfo {author} {\bibfnamefont {N.~E.}\ \bibnamefont
  {Sluchanko}},\ }\href {https://doi.org/10.1134/S1063776107070278} {\bibfield
  {journal} {\bibinfo  {journal} {J. Exp. Theor. Phys.}\ }\textbf {\bibinfo
  {volume} {105}},\ \bibinfo {pages} {132} (\bibinfo {year}
  {2007})}\BibitemShut {NoStop}%
\bibitem [{\citenamefont {Midya}\ \emph {et~al.}(2019)\citenamefont {Midya},
  \citenamefont {Rubi},\ and\ \citenamefont
  {Mahendiran}}]{Midya2019Colossalmagnetoresistanceinlow-doped}%
  \BibitemOpen
  \bibfield  {author} {\bibinfo {author} {\bibfnamefont {A.}~\bibnamefont
  {Midya}}, \bibinfo {author} {\bibfnamefont {K.}~\bibnamefont {Rubi}},\ and\
  \bibinfo {author} {\bibfnamefont {R.}~\bibnamefont {Mahendiran}},\ }\href
  {https://doi.org/10.1063/1.5071448} {\bibfield  {journal} {\bibinfo
  {journal} {J. Appl. Phys.}\ }\textbf {\bibinfo {volume} {125}},\ \bibinfo
  {pages} {023910} (\bibinfo {year} {2019})}\BibitemShut {NoStop}%
\bibitem [{\citenamefont {Wang}\ \emph {et~al.}(2022)\citenamefont {Wang},
  \citenamefont {Been}, \citenamefont {Gaudet}, \citenamefont {Alqasseri},
  \citenamefont {Fruhling}, \citenamefont {Yao}, \citenamefont {Stuhr},
  \citenamefont {Zhu}, \citenamefont {Ren}, \citenamefont {Cui}, \citenamefont
  {Jia}, \citenamefont {Moritz}, \citenamefont {Chowdhury}, \citenamefont
  {Devereaux},\ and\ \citenamefont
  {Tafti}}]{Wang2022Anisotropyofthemagneticandtransportproperties}%
  \BibitemOpen
  \bibfield  {author} {\bibinfo {author} {\bibfnamefont {Z.-C.}\ \bibnamefont
  {Wang}}, \bibinfo {author} {\bibfnamefont {E.}~\bibnamefont {Been}}, \bibinfo
  {author} {\bibfnamefont {J.}~\bibnamefont {Gaudet}}, \bibinfo {author}
  {\bibfnamefont {G.~M.~A.}\ \bibnamefont {Alqasseri}}, \bibinfo {author}
  {\bibfnamefont {K.}~\bibnamefont {Fruhling}}, \bibinfo {author}
  {\bibfnamefont {X.}~\bibnamefont {Yao}}, \bibinfo {author} {\bibfnamefont
  {U.}~\bibnamefont {Stuhr}}, \bibinfo {author} {\bibfnamefont
  {Q.}~\bibnamefont {Zhu}}, \bibinfo {author} {\bibfnamefont {Z.}~\bibnamefont
  {Ren}}, \bibinfo {author} {\bibfnamefont {Y.}~\bibnamefont {Cui}}, \bibinfo
  {author} {\bibfnamefont {C.}~\bibnamefont {Jia}}, \bibinfo {author}
  {\bibfnamefont {B.}~\bibnamefont {Moritz}}, \bibinfo {author} {\bibfnamefont
  {S.}~\bibnamefont {Chowdhury}}, \bibinfo {author} {\bibfnamefont
  {T.}~\bibnamefont {Devereaux}},\ and\ \bibinfo {author} {\bibfnamefont
  {F.}~\bibnamefont {Tafti}},\ }\href
  {https://doi.org/10.1103/PhysRevB.105.165122} {\bibfield  {journal} {\bibinfo
   {journal} {Phys. Rev. B}\ }\textbf {\bibinfo {volume} {105}},\ \bibinfo
  {pages} {165122} (\bibinfo {year} {2022})}\BibitemShut {NoStop}%
\bibitem [{\citenamefont {Blawat}\ \emph {et~al.}(2022)\citenamefont {Blawat},
  \citenamefont {Marshall}, \citenamefont {Singleton}, \citenamefont {Feng},
  \citenamefont {Cao}, \citenamefont {Xie},\ and\ \citenamefont
  {Jin}}]{Blawat2022UnusualElectricalandMagnetic}%
  \BibitemOpen
  \bibfield  {author} {\bibinfo {author} {\bibfnamefont {J.}~\bibnamefont
  {Blawat}}, \bibinfo {author} {\bibfnamefont {M.}~\bibnamefont {Marshall}},
  \bibinfo {author} {\bibfnamefont {J.}~\bibnamefont {Singleton}}, \bibinfo
  {author} {\bibfnamefont {E.}~\bibnamefont {Feng}}, \bibinfo {author}
  {\bibfnamefont {H.}~\bibnamefont {Cao}}, \bibinfo {author} {\bibfnamefont
  {W.}~\bibnamefont {Xie}},\ and\ \bibinfo {author} {\bibfnamefont
  {R.}~\bibnamefont {Jin}},\ }\href
  {https://doi.org/https://doi.org/10.1002/qute.202200012} {\bibfield
  {journal} {\bibinfo  {journal} {Adv. Quantum Technol.}\ }\textbf {\bibinfo
  {volume} {5}},\ \bibinfo {pages} {2200012} (\bibinfo {year}
  {2022})}\BibitemShut {NoStop}%
\bibitem [{\citenamefont {Souza}\ \emph {et~al.}(2022)\citenamefont {Souza},
  \citenamefont {Thomas}, \citenamefont {Bauer}, \citenamefont {Thompson},
  \citenamefont {Ronning}, \citenamefont {Pagliuso},\ and\ \citenamefont
  {Rosa}}]{souza2022}%
  \BibitemOpen
  \bibfield  {author} {\bibinfo {author} {\bibfnamefont {J.~C.}\ \bibnamefont
  {Souza}}, \bibinfo {author} {\bibfnamefont {S.~M.}\ \bibnamefont {Thomas}},
  \bibinfo {author} {\bibfnamefont {E.~D.}\ \bibnamefont {Bauer}}, \bibinfo
  {author} {\bibfnamefont {J.~D.}\ \bibnamefont {Thompson}}, \bibinfo {author}
  {\bibfnamefont {F.}~\bibnamefont {Ronning}}, \bibinfo {author} {\bibfnamefont
  {P.~G.}\ \bibnamefont {Pagliuso}},\ and\ \bibinfo {author} {\bibfnamefont
  {P.~F.~S.}\ \bibnamefont {Rosa}},\ }\href
  {https://doi.org/10.1103/PhysRevB.105.035135} {\bibfield  {journal} {\bibinfo
   {journal} {Phys. Rev. B}\ }\textbf {\bibinfo {volume} {105}},\ \bibinfo
  {pages} {035135} (\bibinfo {year} {2022})}\BibitemShut {NoStop}%
\bibitem [{\citenamefont {Gati}\ \emph {et~al.}(2021)\citenamefont {Gati},
  \citenamefont {Bud'ko}, \citenamefont {Wang}, \citenamefont {Valadkhani},
  \citenamefont {Gupta}, \citenamefont {Kuthanazhi}, \citenamefont {Xiang},
  \citenamefont {Wilde}, \citenamefont {Sapkota}, \citenamefont {Guguchia},
  \citenamefont {Khasanov}, \citenamefont {Valent\'{\i}},\ and\ \citenamefont
  {Canfield}}]{Gati2021Pressureinducedferromagnetisminthetopological}%
  \BibitemOpen
  \bibfield  {author} {\bibinfo {author} {\bibfnamefont {E.}~\bibnamefont
  {Gati}}, \bibinfo {author} {\bibfnamefont {S.~L.}\ \bibnamefont {Bud'ko}},
  \bibinfo {author} {\bibfnamefont {L.-L.}\ \bibnamefont {Wang}}, \bibinfo
  {author} {\bibfnamefont {A.}~\bibnamefont {Valadkhani}}, \bibinfo {author}
  {\bibfnamefont {R.}~\bibnamefont {Gupta}}, \bibinfo {author} {\bibfnamefont
  {B.}~\bibnamefont {Kuthanazhi}}, \bibinfo {author} {\bibfnamefont
  {L.}~\bibnamefont {Xiang}}, \bibinfo {author} {\bibfnamefont {J.~M.}\
  \bibnamefont {Wilde}}, \bibinfo {author} {\bibfnamefont {A.}~\bibnamefont
  {Sapkota}}, \bibinfo {author} {\bibfnamefont {Z.}~\bibnamefont {Guguchia}},
  \bibinfo {author} {\bibfnamefont {R.}~\bibnamefont {Khasanov}}, \bibinfo
  {author} {\bibfnamefont {R.}~\bibnamefont {Valent\'{\i}}},\ and\ \bibinfo
  {author} {\bibfnamefont {P.~C.}\ \bibnamefont {Canfield}},\ }\href
  {https://doi.org/10.1103/PhysRevB.104.155124} {\bibfield  {journal} {\bibinfo
   {journal} {Phys. Rev. B}\ }\textbf {\bibinfo {volume} {104}},\ \bibinfo
  {pages} {155124} (\bibinfo {year} {2021})}\BibitemShut {NoStop}%
\bibitem [{\citenamefont {Jose}\ \emph {et~al.}(2023)\citenamefont {Jose},
  \citenamefont {Burrage}, \citenamefont {Jimenez}, \citenamefont {Xie},
  \citenamefont {Lavina}, \citenamefont {Zhao}, \citenamefont {Alp},
  \citenamefont {Zhang}, \citenamefont {Xiao}, \citenamefont {Vohra},\ and\
  \citenamefont {Bi}}]{joseEvolutionMagnetismValence2023}%
  \BibitemOpen
  \bibfield  {author} {\bibinfo {author} {\bibfnamefont {G.~C.}\ \bibnamefont
  {Jose}}, \bibinfo {author} {\bibfnamefont {K.}~\bibnamefont {Burrage}},
  \bibinfo {author} {\bibfnamefont {J.~L.~G.}\ \bibnamefont {Jimenez}},
  \bibinfo {author} {\bibfnamefont {W.}~\bibnamefont {Xie}}, \bibinfo {author}
  {\bibfnamefont {B.}~\bibnamefont {Lavina}}, \bibinfo {author} {\bibfnamefont
  {J.}~\bibnamefont {Zhao}}, \bibinfo {author} {\bibfnamefont {E.~E.}\
  \bibnamefont {Alp}}, \bibinfo {author} {\bibfnamefont {D.}~\bibnamefont
  {Zhang}}, \bibinfo {author} {\bibfnamefont {Y.}~\bibnamefont {Xiao}},
  \bibinfo {author} {\bibfnamefont {Y.~K.}\ \bibnamefont {Vohra}},\ and\
  \bibinfo {author} {\bibfnamefont {W.}~\bibnamefont {Bi}},\ }\href
  {https://doi.org/10.1103/PhysRevB.107.245121} {\bibfield  {journal} {\bibinfo
   {journal} {Phys. Rev. B}\ }\textbf {\bibinfo {volume} {107}},\ \bibinfo
  {pages} {245121} (\bibinfo {year} {2023})}\BibitemShut {NoStop}%
\bibitem [{\citenamefont {Park}\ \emph {et~al.}(2002)\citenamefont {Park},
  \citenamefont {Eun}, \citenamefont {Kang},\ and\ \citenamefont
  {Kim}}]{Park2002Eu5In2Sb6Eu5In2xZnxSb6}%
  \BibitemOpen
  \bibfield  {author} {\bibinfo {author} {\bibfnamefont {S.}~\bibnamefont
  {Park}}, \bibinfo {author} {\bibfnamefont {S.}~\bibnamefont {Eun}}, \bibinfo
  {author} {\bibfnamefont {W.}~\bibnamefont {Kang}},\ and\ \bibinfo {author}
  {\bibfnamefont {S.}~\bibnamefont {Kim}},\ }\href
  {https://doi.org/10.1039/b106812a} {\bibfield  {journal} {\bibinfo  {journal}
  {J. Mater. Chem.}\ }\textbf {\bibinfo {volume} {12}},\ \bibinfo {pages}
  {1839} (\bibinfo {year} {2002})}\BibitemShut {NoStop}%
\end{thebibliography}%
\bibliographystyle{apsrev4-2}

\end{document}


\large

\setcounter{figure}{0}
\setcounter{table}{0}
\setcounter{equation}{0}
\setcounter{section}{0}
\setcounter{page}{1}
	

\title{Supplemental Materials: Colossal magnetoresistance in narrow-gap magnetic semiconductor Eu$_2$CdAs$_2$}

\author{Zhiyu Zhou}
\altaffiliation{Contributed equally to this work}
\affiliation{Key Laboratory of Quantum Materials and Devices of Ministry of Education, School of Physics, Southeast University, Nanjing 211189, China}
\author{Zhi-Cheng Wang}
\altaffiliation{Contributed equally to this work}
\email{wzc@seu.edu.cn}
\affiliation{Key Laboratory of Quantum Materials and Devices of Ministry of Education, School of Physics, Southeast University, Nanjing 211189, China}
\author{Xiyu Chen}
\affiliation{Key Laboratory of Quantum Materials and Devices of Ministry of Education, School of Physics, Southeast University, Nanjing 211189, China}
\author{Xin-Yu Zhao}
\affiliation{School of Physics, Interdisciplinary Center for Quantum Information and State Key Laboratory of Silicon and Advanced Semiconductor Materials, Zhejiang University, Hangzhou 310058, China}
\author{Yaran Duan}
\affiliation{Key Laboratory of Quantum Materials and Devices of Ministry of Education, School of Physics, Southeast University, Nanjing 211189, China}
\author{Shiran Tao}
\affiliation{Key Laboratory of Quantum Materials and Devices of Ministry of Education, School of Physics, Southeast University, Nanjing 211189, China}
\author{Guang-Han Cao}
\affiliation{School of Physics, Interdisciplinary Center for Quantum Information and State Key Laboratory of Silicon and Advanced Semiconductor Materials, Zhejiang University, Hangzhou 310058, China}

\author{Shuai Dong}
\affiliation{Key Laboratory of Quantum Materials and Devices of Ministry of Education, School of Physics, Southeast University, Nanjing 211189, China}

\date{\today}

\maketitle

	\begin{center}
		\textbf{Contents}
		\begin{enumerate}
			\item Figure S1: Orientation photograph of a single crystal.
			\item Table S1: Crystallographic data and refinement results.
			\item Figure S2: Magnetic properties of Eu$_2$CdAs$_2$.
			\item Figure S3: Specific heat capacity analysis.
			\item Figure S4: Magnetotransport with $I \parallel a$ and $H \parallel a$.
			\item Table S2: Fitting details and parameters for low-temperature activation energies.
		\end{enumerate}
	\end{center}
	\clearpage
	
	\subsection{Orientation Photograph of a Single Crystal}
	
	\begin{figure}[h]
		\centering
		\includegraphics[width=\columnwidth]{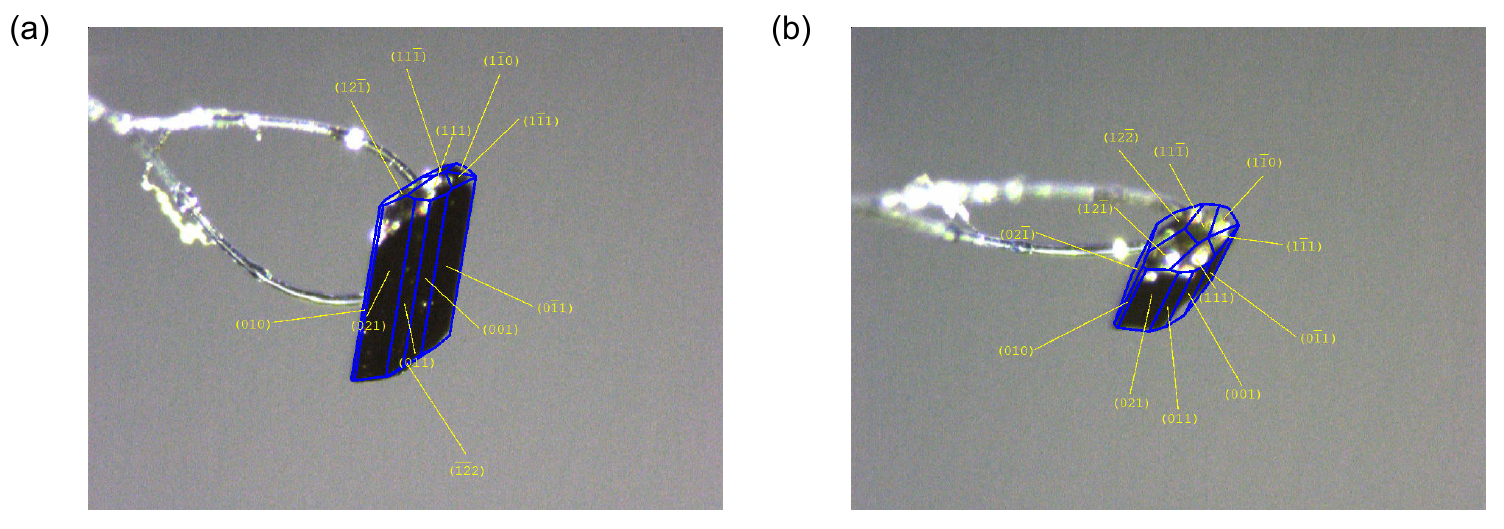}
		\caption{Orientation photograph of a Eu$_2$CdAs$_2$ single crystal. The Miller indices of each crystal facet are labeled in yellow text with blue boxes, indicating that the long growth axis of the crystal corresponds to the crystallographic $a$-axis.}
		\label{fig:FigS1}
	\end{figure}
	
	The crystallographic orientation of the Eu$_2$CdAs$_2$ single crystal was determined by single-crystal X-ray diffraction (SCXRD). The results demonstrate that the long axis of crystal growth corresponds directly to the crystallographic $a$-axis.
	
	\clearpage
	
	\subsection{Crystallographic Data}
	
	\begin{table}[htbp]
		\centering
		\caption{Crystallographic data and refinement results for Eu$_2$CdAs$_2$ obtained from SCXRD at 150 K.}
		\begin{tabular}{ll}
			\hline
			Material & Eu$_2$CdAs$_2$ \\
			\hline
			Crystal system  & Orthorhombic \\	
			Space group & $Cmc2_1$ (No. 36) \\   
			$a$ (\AA) & 4.4182(9) \\                               
			$b$ (\AA) & 16.764(3) \\
			$c$ (\AA) & 7.4527(17) \\
			$V$ (\AA$^3$) & 552.0(2) \\                           
			$Z$ & 4 \\
			$\rho_\mathrm{calc}$ (g/cm$^3$) & 6.813 \\
			Temperature (K) & 150 \\
			Radiation & Mo $K\alpha$ \\
			Reflections collected & 2524 \\
			Independent reflections & 571 \\
			$R_{\mathrm{int}}$ & 0.0293 \\
			Goodness-of-fit & 0.910 \\
			$R_1^{\text{a)}}$ & 0.0113 \\
			$wR_2^{\text{b)}}$ & 0.0235 \\
			\hline
		\end{tabular}
		
		\vspace{2mm}
		\leftline{\footnotesize $^{\text{a)}} R_1 = \Sigma||F_o|-|F_c||/\Sigma|F_o|$; $^{\text{b)}} wR_2 = [\Sigma w(F_o^2-F_c^2)^2/\Sigma w(F_o^2)^2]^{1/2}$.}
		\label{tab:S1}
	\end{table}
	
	The crystallographic data obtained from SCXRD confirm that the crystal structure of Eu$_2$CdAs$_2$ is in excellent agreement with previously reported literature (Inorg. Chem. \textbf{50}, 8020 (2011)).
	
	\clearpage
	
	\subsection{Magnetic Data of Eu$_2$CdAs$_2$}
	
	\begin{figure}[h]
		\centering
		\includegraphics[width=\columnwidth]{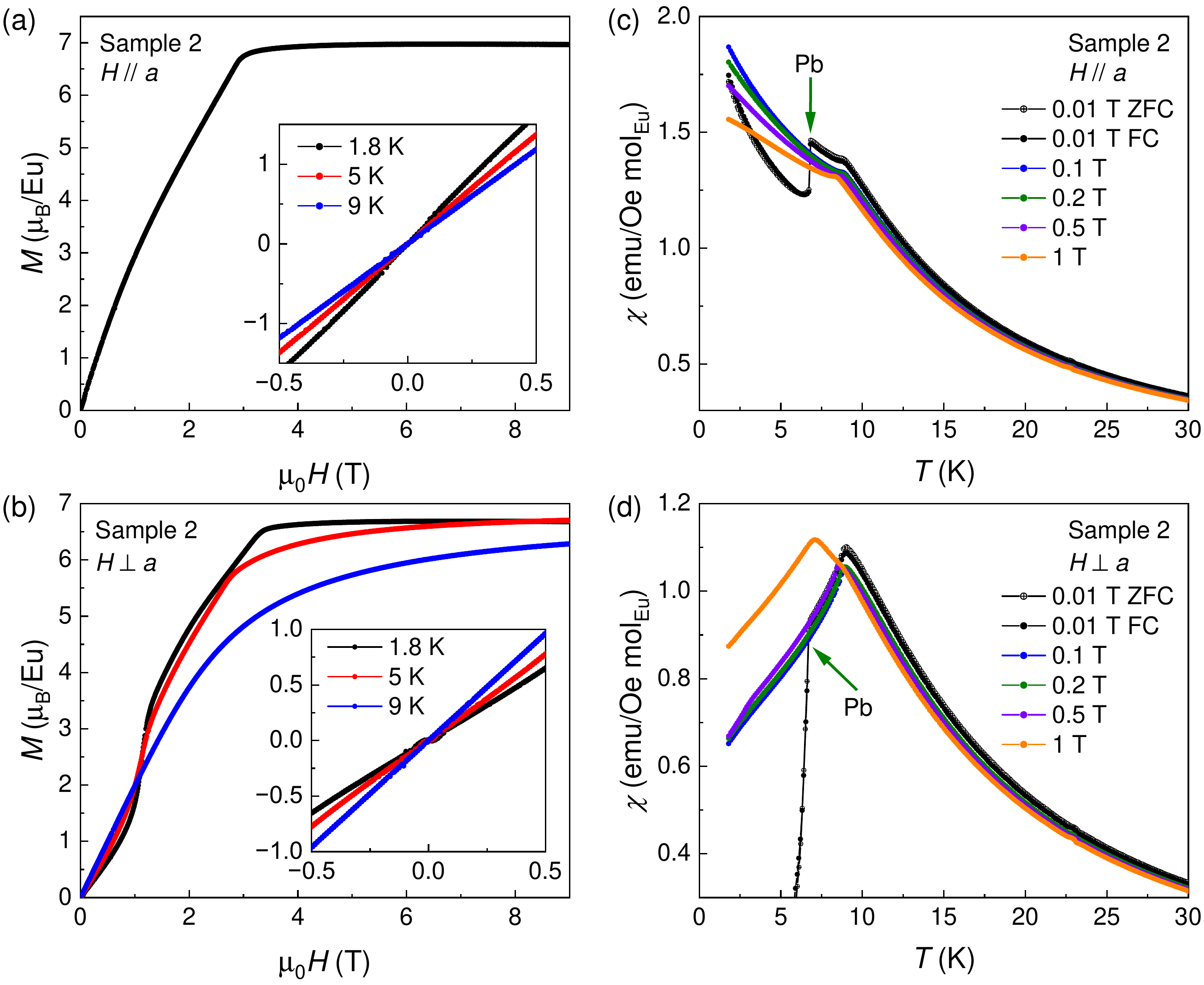}
		\caption{Magnetic characterization of Eu$_2$CdAs$_2$ (Sample 2). (a, b) Magnetization as a function of the applied magnetic field (a) parallel to ($H \parallel a$) and (b) perpendicular to ($H \perp a$) the $a$-axis; the insets show the corresponding low-field behaviors. (c, d) Temperature-dependent magnetic susceptibility under various magnetic fields applied (c) parallel to and (d) perpendicular to the $a$-axis. These data demonstrate that no magnetic hysteresis loop is observed, confirming that the magnetic properties are highly reproducible across different sample batches.}
		\label{fig:FigS2}
	\end{figure}
	
	Figure \ref{fig:FigS2} displays the temperature-dependent magnetic susceptibility and magnetization curves of a different Eu$_2$CdAs$_2$ single crystal (Sample 2), which is complementary to Sample 1 discussed in the main text. The magnetic behaviors of both Sample 1 and Sample 2 consistently indicate the absence of a macroscopic net ferromagnetic component in Eu$_2$CdAs$_2$, demonstrating that the magnetic properties are highly reproducible across different crystal growth batches.
	
	\clearpage
	
	\subsection{Analysis of Heat Capacity}
	
	\begin{figure}[h]
		\centering
		\includegraphics[width=\columnwidth]{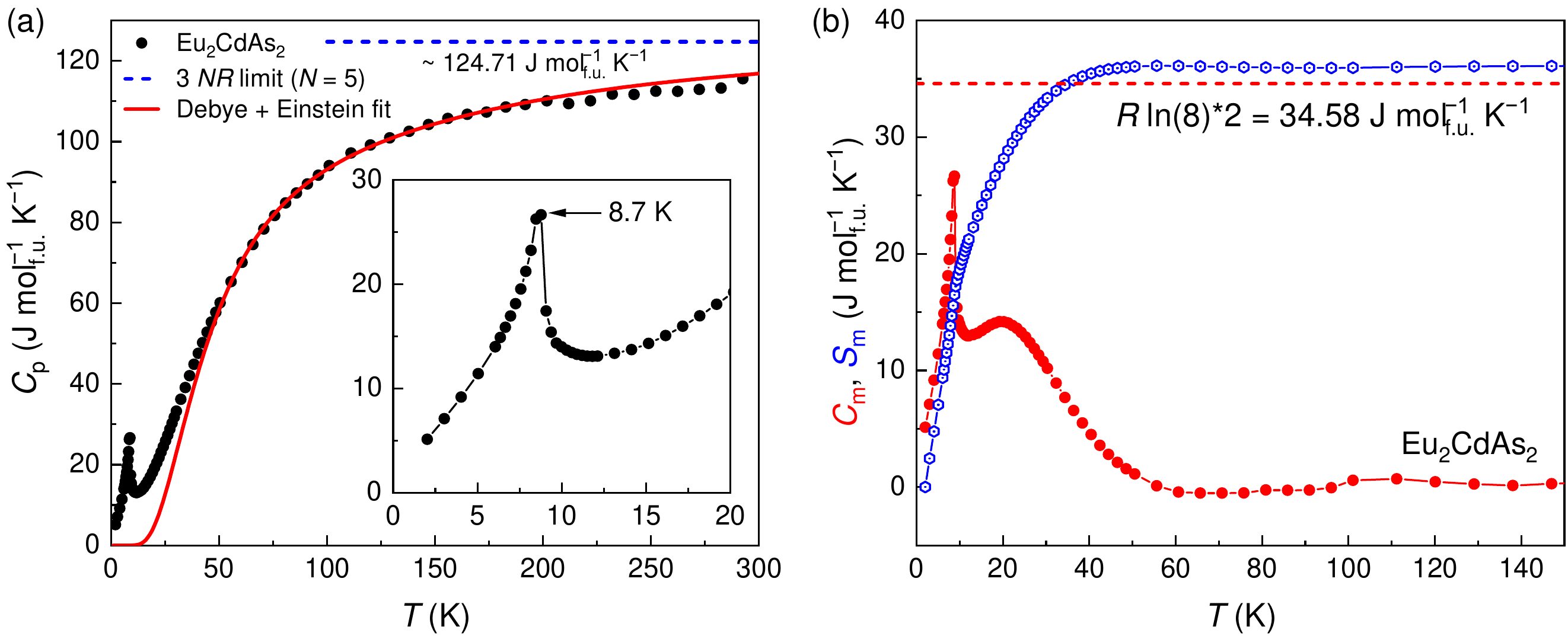}
		\caption{(a) Temperature dependence of the zero-field specific heat capacity, $C_\mathrm{p}(T)$, fitted from 50 to 200 K using a combination of the Debye and Einstein models (red curve). The inset shows a magnified view of the $C_\mathrm{p}(T)$ data below 20 K, highlighting the phase transition at 8.7 K. (b) Magnetic heat capacity $C_\mathrm{m}(T)$ (red dots) and magnetic entropy $S_\mathrm{m}(T)$ (blue circles) derived from $C_\mathrm{m}(T)$. The red dashed line indicates the expected theoretical magnetic entropy for two Eu$^{2+}$ ions.}
		\label{fig:FigS3}
	\end{figure}
	
	Figure \ref{fig:FigS3} presents the specific heat capacity ($C_\mathrm{p}(T)$) of Eu$_2$CdAs$_2$ (Sample 2). As depicted in Fig. \ref{fig:FigS3}(a), the room-temperature $C_\mathrm{p}$ value falls slightly below the Dulong-Petit limit of 124.71 J mol$^{-1}$ K$^{-1}$, which represents the theoretical limit for Eu$_2$CdAs$_2$. The lower-right inset reveals a distinct $\lambda$-type peak at 8.7 K, which is consistent with the Néel temperature $T_\mathrm{N} = 9$ K determined from magnetic susceptibility measurements, thereby confirming the bulk nature of the antiferromagnetic transition.
	
	The zero-field $C_\mathrm{p}(T)$ data within the temperature range of 50--200 K were analyzed by combining the Debye and Einstein models according to the formula:
	
	\begin{equation}
		C_\mathrm{p}(T) = \alpha C_\mathrm{D}(T) + (1 - \alpha) C_\mathrm{E}(T),
		\label{eq:1}
	\end{equation}
	
	The lattice contributions from the two models are expressed respectively as:
	
	\begin{equation}
		C_\mathrm{D}(T) = 9NR \left( \frac{T}{\Theta_\mathrm{D}} \right)^3 \int_{0}^{\Theta_\mathrm{D}/T} \frac{x^4 e^x}{(e^x - 1)^2} \, dx,
		\label{eq:2}
	\end{equation}
	
	\begin{equation}
		C_\mathrm{E}(T) = 3NR \left( \frac{\Theta_\mathrm{E}}{T} \right)^2 \frac{e^{\Theta_\mathrm{E}/T}}{\left( e^{\Theta_\mathrm{E}/T} - 1 \right)^2}.
		\label{eq:3}
	\end{equation}
	
	Here, $C_\mathrm{D}(T)$ and $C_\mathrm{E}(T)$ represent the lattice contributions from the Debye and Einstein models, respectively. $\Theta_\mathrm{D}$ and $\Theta_\mathrm{E}$ denote the Debye and Einstein temperatures, respectively, and $\alpha$ determines the relative weights of the Debye and Einstein components. The fitted parameters are determined to be $\alpha = 0.139$, $\Theta_\mathrm{D} = 911.15$ K, and $\Theta_\mathrm{E} = 140.07$ K.
	
	Using these fitted parameters, we extracted the magnetic contribution to the specific heat ($C_\mathrm{m}$) by subtracting the simulated lattice component. The magnetic entropy ($S_\mathrm{m}$) was subsequently calculated via numerical integration, and the corresponding results are plotted in Fig. \ref{fig:FigS3}(b). The saturated $S_\mathrm{m}$ value reaches 36.1 J mol$^{-1}$ K$^{-1}$, which is very close to the theoretical prediction of $S_\mathrm{m} = 2R \ln(2S + 1)$ for $S = 7/2$ (34.58 J mol$^{-1}$ K$^{-1}$). 
	
	\clearpage
	
	\subsection{Magnetotransport with $I \parallel a$ and $H \parallel a$}
	
	\begin{figure}[htbp]
		\centering
		\includegraphics[width=0.5\columnwidth]{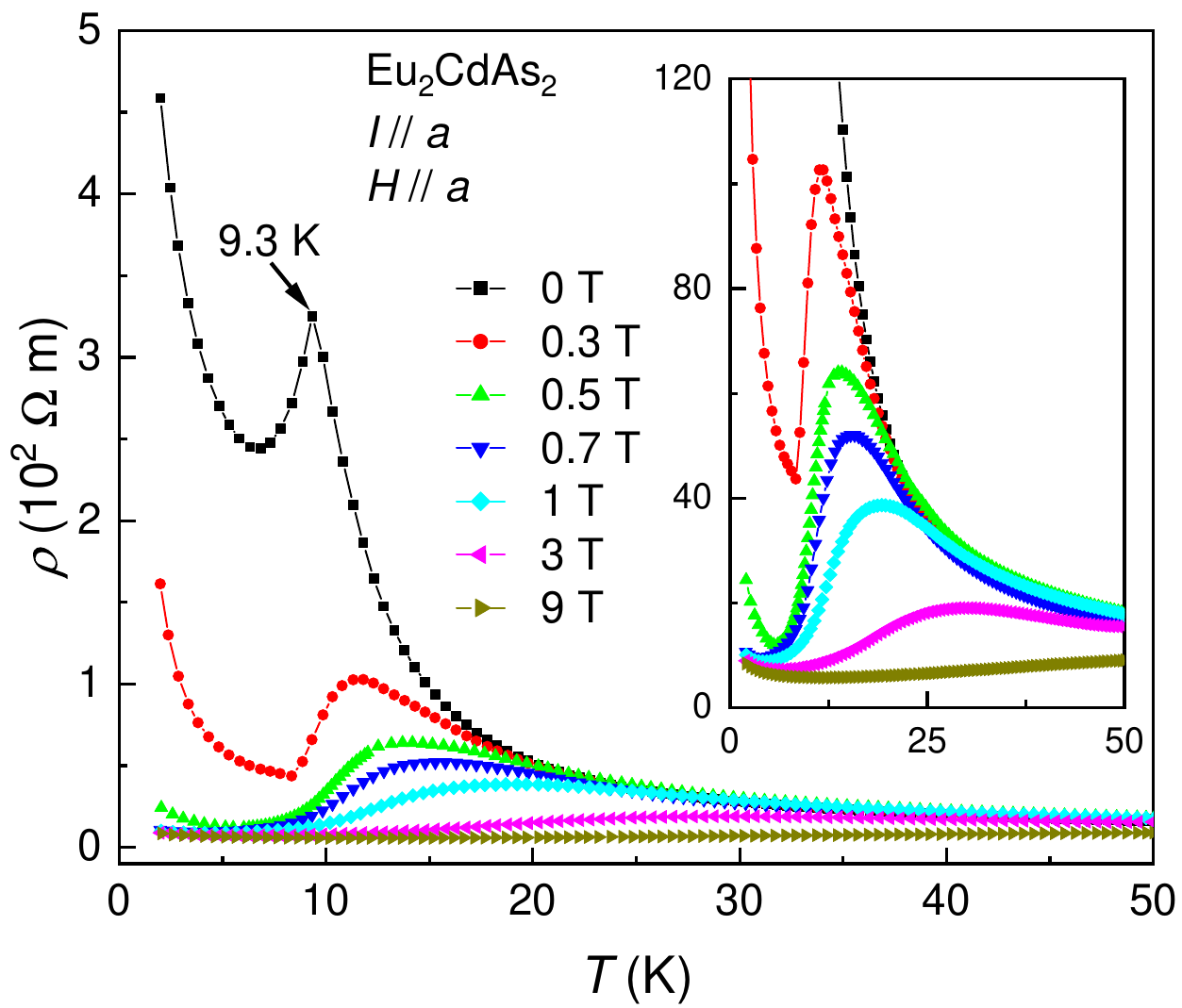}
		\caption{Temperature dependence of the resistivity under various applied magnetic fields parallel to the $a$-axis.}
		\label{fig:FigS4}
	\end{figure}
	
	Figure \ref{fig:FigS4} illustrates the temperature-dependent resistivity curves for the configuration where both the electrical current and the applied magnetic field are parallel to the $a$-axis ($H \parallel a$ and $I \parallel a$). The resulting behavior is highly consistent with the temperature-dependent resistivity obtained under the $H \perp a$ and $I \parallel a$ configuration shown in Fig. 3(b) of the main text. This strong agreement indicates that changing the orientation of the magnetic field has a negligible effect on the overall magnitude of the magnetoresistance.
	
	\clearpage
	
	\subsection{Activation Energies of Reference Materials}
	
	\begin{table}[htbp]
		\centering
		\small
		\caption{Low-temperature activation energies ($E_\mathrm{a}$) of the Eu-based magnetoresistive materials compiled in Fig.~4 of the main text. All $E_\mathrm{a}$ values were extracted by fitting the low-temperature transport data near the maximum of $-\mathrm{MR}$ using the same Arrhenius thermal activation model. The $E_\mathrm{a}$ values and fitting ranges reported in the corresponding literature are listed for comparison.}
		\label{tab:S2}
		\resizebox{\textwidth}{!}{%
		\begin{tabular}{l |c |c| c| l| l}
		\hline
		Material & $E_\mathrm{a}$ (meV) & Error (\%) & Range (K) & Reference & Literature $E_\mathrm{a}$ (meV) \& range (K) \\
		\hline
		EuB$_6$ & 0.395 $\pm$ 0.017 & 4.20 & 19--30 & J. Exp. Theor. Phys. 105, 132 (2007) &  \\
		Eu$_{0.74}$Ca$_{0.26}$B$_6$ & 11.57 $\pm$ 0.30 & 2.59 & 15.5--24.5 & J. Exp. Theor. Phys. 111, 246 (2010) &  \\
		EuSe$_2$ & 25.40 $\pm$ 0.65 & 2.32 & 15.5--20.5 & Phys. Rev. B 112, L140405 (2025) & 32 meV, 14--23 K \\
		EuTe$_2$ & 0.55 $\pm$ 0.01 & 1.65 & 1.8--2.8 & Phys. Rev. Mater. 4, 013405 (2020) & 16.24 meV, 30--300 K \\
		EuMnSb$_2$ (1) & 8.36 $\pm$ 0.65 & 7.78 & 34--47 & npj Quantum Mater. 6, 94 (2021) & 37.95 meV, 200--300 K \\
		EuMnSb$_2$ (2) & 6.13 $\pm$ 0.32 & 5.23 & 31.5--50 & Phys. Rev. B 96, 205103 (2017) & 30.1 meV, 150--300 K \\
		Eu$_{11}$Zn$_4$Sn$_2$As$_{12}$ & 12.97 $\pm$ 0.49 & 3.81 & 15--16 & Chem. Mater. 30, 7067 (2018) & 19.8 meV, 70--170 K \\
		Eu$_5$In$_2$As$_6$ & 30.97 $\pm$ 0.31 & 1.00 & 29--54 & Phys. Rev. B 111, 115114 (2025) &  \\
		Eu$_5$In$_2$Sb$_6$ & 6.46 $\pm$ 0.17 & 2.70 & 17--26 & npj Quantum Mater. 5, 52 (2020) & 42 meV, 50--150 K \\
		Eu$_5$Sn$_2$As$_6$ & 28.77 $\pm$ 0.40 & 1.39 & 34--41.5 & Phys. Rev. B 111, 054406 (2025) & 35 meV, 30--50 K \\
		EuTiO$_3$ & 12.91 $\pm$ 0.14 & 1.06 & 20--32 & Eur. Phys. J. B 91, 36 (2018) & 13 meV, 20--40 K \\
		Eu$_{0.99}$La$_{0.01}$TiO$_3$ & 0 & 0 & Metallic & EPL 118, 57008 (2017) &  \\
		EuTi$_{1-x}$Nb$_x$O$_3$ & 0.8 $\pm$ 0.1 & 12.9 & 23--37 & J. Appl. Phys. 125, 023910 (2019) &  \\
		EuZn$_2$P$_2$ (1) & 15.3 $\pm$ 0.7 & 4.24 & 46--53 & Phys. Rev. B 108, 045116 (2023) &  \\
		EuZn$_2$P$_2$ (2) & 22.09 $\pm$ 0.23 & 1.04 & 25--45 & Phys. Rev. Mater. 9, 104403 (2025) & S1: 20 meV, 30--200 K; S2: 173 meV, 100--300 K \\
		EuZn$_2$As$_2$ (1) & 2.45 $\pm$ 0.023 & 0.934 & 19--38 & Phys. Rev. B 105, 165122 (2022) &  \\
		EuZn$_2$As$_2$ (2) & 3.06 $\pm$ 0.04 & 1.35 & 21--48 & Phys. Rev. B 108, 205140 (2023) &  \\
		EuZn$_2$As$_2$ (3) & 3.76 $\pm$ 0.13 & 3.44 & 27--35.1 & Adv. Quantum Technol. 5, 2200012 (2022) &  \\
		EuCd$_2$P$_2$ (1) & 13.02 $\pm$ 0.43 & 3.30 & 19--23 & Adv. Mater. 33, 2005755 (2021) &  \\
		EuCd$_2$P$_2$ (2) & 11.37 $\pm$ 0.25 & 2.18 & 19--23.5 & Phys. Rev. B 109, 104421 (2024) &  \\
		EuCd$_2$As$_2$ (1) & 0.825 $\pm$ 0.033 & 4.00 & 31--44 & Adv. Sci. 10, 2207121 (2023) &  \\
		EuCd$_2$As$_2$ (2) & Adopted from literature &  &  & Phys. Rev. B 109, 125202 (2024) & 27.9 meV, 30--100 K \\
		EuCd$_2$Sb$_2$ & 0.0053 $\pm$ 0.0006 & 11.8 & 9--15.5 & Phys. Rev. B 98, 064419 (2018) &  \\
		\hline
		\end{tabular}}
	\end{table}
	
	\clearpage